\newcommand {\source} {MACS\,J0358.8--2955}

\documentclass[usenatbib,useAMS]{mn2e}

\usepackage{graphicx,times}
\usepackage{longtable,deluxetable}

\title[The 3D geometry of MACS\,J0358.8--2955] {The three-dimensional geometry and merger
  history of the massive galaxy cluster MACS\,J0358.8--2955}

\author[Hsu, Ebeling, Richard] {Li-Yen Hsu$^1$, Harald Ebeling$^1$,
  Johan Richard$^2$\\ $^1$Institute for Astronomy, University of
  Hawaii, 2680 Woodlawn Drive, Honolulu, HI 96822, USA\\ $^2$Centre de
  Recherche Astrophysique de Lyon, Universit\'e Lyon 1, 9 Avenue
  Charles Andr\'e, 69561 Saint Genis Laval Cedex, France}

\begin{document}

\label{firstpage}

\maketitle

\begin{abstract}
We present results of a combined X-ray/optical analysis of the
dynamics of the massive cluster \source\ ($z=0.428$) based on
observations with the Chandra X-ray Observatory, the Hubble Space
Telescope, and the Keck-I telescope on Mauna Kea. \source\ is found to
be one of the most X-ray luminous clusters known at $z{>}0.3$,
featuring $L_{X,\mathrm{bol}}(<r_{500})=4.24 \times 10^{45}$ erg
s$^{-1}$, kT = $9.55^{+0.58}_{-0.37}$ keV, $M^{\rm 3D}_{\rm
  gas}(<r_{500}) = (9.18 \pm 1.45) \times 10^{13} M_{\odot}$, and
$M^{\rm 3D}_{tot}(<r_{500}) = (1.12 \pm 0.18) \times 10^{15}
M_{\odot}$.  The system's high velocity dispersion of
$1440^{+130}_{-110}\rm~km~s^{-1}$ (890 km s$^{-1}$ when the correct relativistic
equation is used), however, is inflated
by infall along the line of sight, as the result of a complex merger
of at least three sub-clusters. One collision proceeds close to
head-on, while the second features a significant impact parameter. The
temperature variations in the intra-cluster gas, two tentative cold
fronts, the radial velocities measured for cluster galaxies, and the
small offsets between collisional and non-collisional cluster
components all suggest that both merger events are observed close to
core passage and along axes that are greatly inclined with respect to
the plane of the sky. A strong-lensing analysis of the system anchored
upon three triple-image systems (two of which have spectroscopic
redshifts) yields independent constraints on the mass
distribution. For a gas fraction of 8.2\%, the resulting
strong-lensing mass profile is in good agreement with our X-ray
estimates, and the details of the mass distribution are fully
consistent with our interpretation of the three-dimensional merger
history of this complex system.

Underlining yet again the power of X-ray selection, our analysis also
resolves earlier confusion about the contribution of the partly
superimposed foreground cluster A\,3192 ($z{=}0.168$). Based on very
faint X-ray emission detected by our Chandra observation and 16
concordant redshifts we identify A\,3192 as two groups of galaxies,
separated by 700 kpc in the plane of the sky.  The X-ray luminosity
and mass of the two components of A\,3192 combined are less than 0.5\%
and less than 8\% of that of \source.

\end{abstract}


\section{Introduction}

Residing and evolving at the nodes of the Cosmic Web
\citep{1996Natur.380..603B}, galaxy clusters represent the latest
stage in the gravity-driven hierarchical growth of structure in the
Universe. The assembly of clusters proceeds both continuously through
the steady accretion of matter from their surroundings as well as
through discrete merger events involving galaxy groups or
clusters. The latter are among the most energetic events in the
Universe and provide us with rare opportunities to exploit the
different collisional behaviour of the cluster components (galaxies,
gas, and dark matter) to study the dynamics and three-dimensional
geometry of the merger. Specifically, head-on collisions of clusters
result in a pronounced segregation of collisional (gas) and
non-collisional matter (galaxies and dark matter) that has been used
to constrain the collisional self-interaction cross section of dark
matter \citep{2004ApJ...606..819M, 2008ApJ...687..959B}. In addition,
the offset between the bulk of the gravitational mass and the bulk of
the viscous mass allows us to differentiate between ram-pressure
stripping and tidal effects on the evolution of galaxies in clusters
(e.g., \citealp{2008ApJ...684..160M, 2011MNRAS.410.2593M}). Finally,
cluster mergers often constitute extremely powerful gravitational
telescopes with very large Einstein radii that permit the study of
distant and otherwise unobservable background galaxies (e.g.,
\citealp{2007ApJ...668..643L,2009ApJ...707L.163S,2011MNRAS.414L..31R}).

In order to identify the most dramatic mergers among truly massive
clusters, a recent study by \citeauthor{2012MNRAS.420.2120M} (2012;
hereafter ME12) inspected the optical and X-ray morphology of a
statistical sample of over 100 highly X-ray luminous clusters at
$z>0.15$. Among the selected extreme mergers are complex and well
studied systems like A\,520
\citep{2000A&A...355..443P,2005ApJ...627..733M,2007ApJ...668..806M,2008A&A...491..379G,2012ApJ...747...96J},
A\,2744
\citep{2004MNRAS.349..385K,2006A&A...449..461B,2011MNRAS.417..333M,2011ApJ...728...27O,2012ApJ...750L..23O},
and MACSJ0717.5+3745
\citep{2003MNRAS.339..913E,2004ApJ...609L..49E,2007ApJ...661L..33E,2009ApJ...693L..56M,2011MNRAS.410.2593M,2009A&A...503..707B,2009A&A...505..991V,2012A&A...544A..71L,Jauzac2012}. While
of great interest for a variety of reasons, these systems do, however,
not match the simple Binary, Head-On Merger (BHOM, ME12) morphology
that is most likely to permit a reconstruction of the merger history
and cluster trajectories. Focusing on BHOM candidates, ME12 lists 17
clusters that meet their X-ray/optical selection criteria. Of the 11
primary candidates, several are again well known\footnote{Among them
  MACSJ0025.4--1222
  \citep{2007ApJ...661L..33E,2008ApJ...687..959B,2010MNRAS.406..121M},
  as well as A\,1758
  \citep{2008PASJ...60..345O,2011A&A...529A..38D,2012ApJ...744...94R}
  and A\,2146 \citep{2010MNRAS.406.1721R,
    2011MNRAS.417L...1R,2012MNRAS.423..236R}.}, while the merger
dynamics of others have only recently been unraveled
(MACSJ0140.0--0555; \citealt*{2012arXiv1207.6235H}). We focus here on
\source, a system from ME12's list of seven secondary BHOM candidates.

Our paper is structured as follows. Section~\ref{sec:source}
introduces  \source, the target of this study. Section~\ref{sec:data} summarises our
observations and data reduction procedures at X-ray and optical
wavelengths. Section~\ref{sec:results} describes our analysis in
detail and presents results. In Section~\ref{sec:discussion} we
discuss the results and their implications for the physics of the
merger, and a summary is given in Section~\ref{sec:summary}. Finally,
we derive and discuss, in an appendix, the properties of the
superimposed foreground cluster A\,3192. Throughout this paper, we
assume the concordance $\Lambda$CDM cosmology with $\rm
H_0=70~km~s^{-1}~Mpc^{-1}$, $\rm\Omega_M=0.27$, and
$\Omega_\Lambda=0.73$, for which 1\arcsec\ corresponds to 5.65\,kpc at
the cluster redshift of $z{=}0.428$ (additional galaxy redshifts
obtained by us for this work -- see Section \ref{sec:redshifts} -- led
to a slight revision of the cluster redshift published by us before).

\section{\source}\label{sec:source}

\begin{figure*}
    \includegraphics[width=17cm]{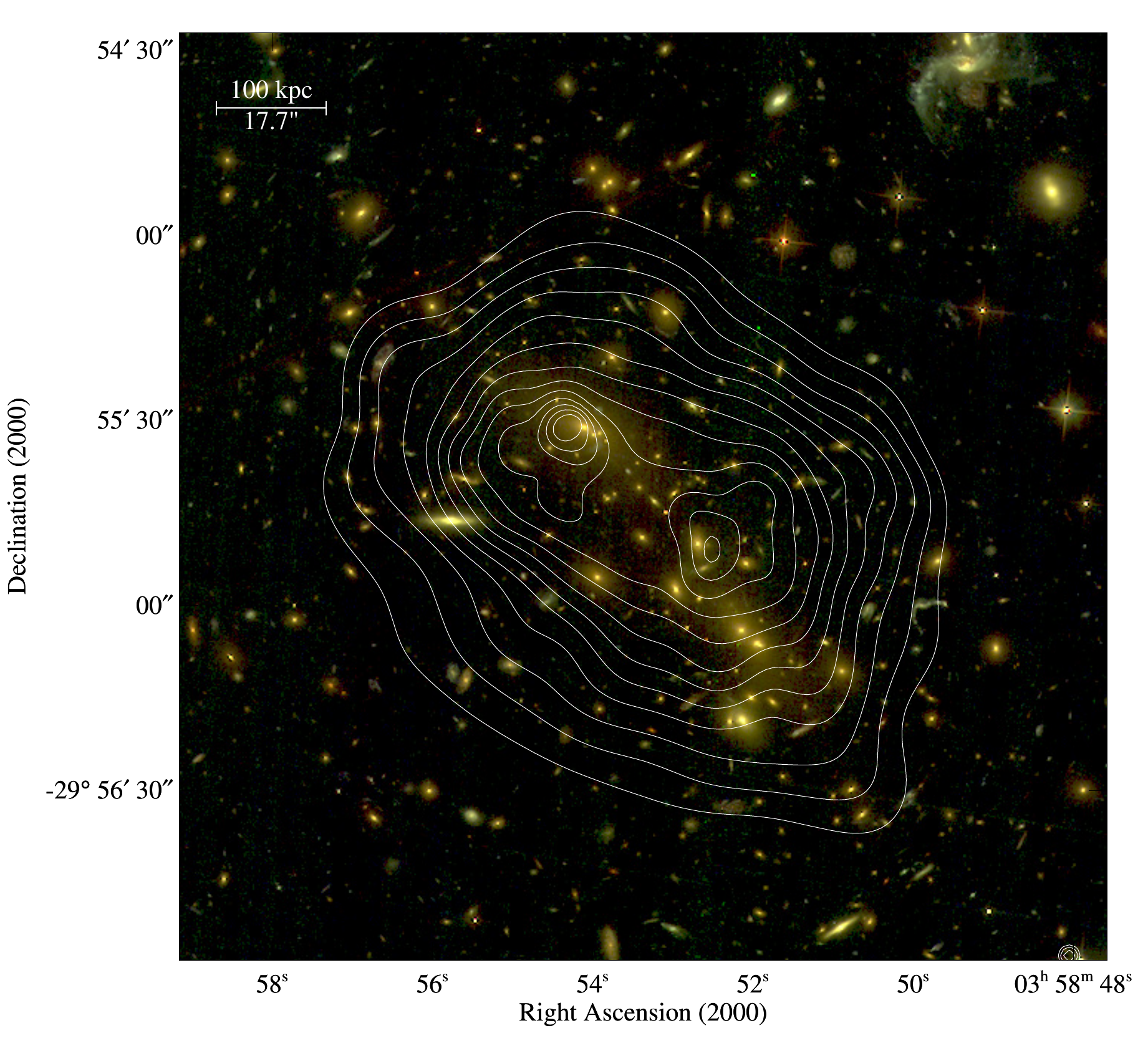}
    \caption{Isointensity contours of the adaptively smoothed
      (ASMOOTH, \citealt{2006MNRAS.368...65E}) X-ray emission from
      \source\ as observed with Chandra (ACIS-I) overlaid on the HST
      colour image (blue: F435W, green: F606W, red: F814W). Contour
      levels are spaced logarithmically.}
    \label{fig:overlay}
\end{figure*}

Based on an X-ray detection in the ROSAT All Sky Survey (RASS,
\citealt{1999A&A...349..389V}) \source\ was discovered in the course
of the Massive Cluster Survey (MACS, \citealt{2001ApJ...553..668E})
and classified as an optically disturbed system at $z=0.425$ based on
13 spectroscopic redshifts \citep{2010MNRAS.407...83E}. The same RASS
X-ray source was previously erroneously identified as A\,3192 (a
nearby foreground cluster at $z=0.168$) by the REFLEX cluster survey
\citep{2004A&A...425..367B}. Owing to the same misidentification, this
presumed nearby cluster was also included in the LoCuSS (e.g.,
\citealt{2008ApJ...682L..73S}) sample of clusters at $z\sim
0.2$. Consistent with our assessment, a recent weak-lensing study
based on an HST snapshot image, combined with groundbased
observations, found \source\ to dominate the mass distribution within
250 kpc of the RASS X-ray position \citep{2012ApJ...748L..23H}.

Figure~\ref{fig:overlay} shows an overlay of the X-ray emission
observed with Chandra onto the optical image of the system obtained by
us with HST (see Section~\ref{sec:data} for observational
details). The foreground system A\,3192 is clearly discernible in the
form of several bright galaxies, mainly north-west of \source; its
contribution to the observed X-ray emission is negligible. We here use
the optical and X-ray data shown in Fig.~\ref{fig:overlay} to
characterise the dynamical state, three-dimensional morphology, and
merger history of all three components of this complex merger (gas,
galaxies, and dark matter).

\section{observation and data reduction}
\label{sec:data}

\subsection{X-ray data}

\source\ was observed three times with the Advanced CCD Imaging
Spectrometer (ACIS, \citealt{2003SPIE.4851...28G}) aboard the Chandra
X-ray Observatory: in October 2009 (9.7 ks; ObsID 11719) and November
2010 (29.7 ks; ObsID 12300, and 20.0 ks; ObsID 13194). All
observations were performed in VFAINT mode. We reduced the data
following standard Chandra data reduction
procedures\footnote{http://cxc.harvard.edu/ciao/index.html} using CIAO
4.4 and CALDB 4.4.8. For the spatial analysis, we merged the three
events files after proper alignment. The spectral fitting, on the
other hand, was performed separately but simultaneously on the three
event files. Point sources were excluded in both spatial and spectral
fits.

In order to avoid instrumental artefacts due to position-dependent
charge transfer inefficiency, we did not measure the background from
an off-source region of our observed datasets, but instead used
appropriately scaled "blank-sky" backgrounds drawn from the
calibration database CALDB. The background
datasets were then merged for the spatial analysis, but dealt with
separately during the spectral fitting.

\subsection{Optical data}

\subsubsection{Hubble Space Telescope imaging}
\label{sec:hstdata}

\source\ was observed on February 19, 2011 in the F435W, F606W and
F814W filters with the Advanced Camera for Surveys (ACS,
\citealt{2004ApJ...600L..93G}) aboard the Hubble Space Telescope for
4500, 2120, and 4620 seconds respectively (GO-12313, PI: Ebeling). At
$z{ =} 0.428$ the field of view of ACS' Wide Field Channel (3.5
$\times$ 3.5 arcmin$^2$) corresponds to approximately 1.2 $\times$ 1.2
Mpc$^2$. Charge Transfer Efficiency (CTE) corrections were applied to
the data using the pixel-based CTE correction code developed by
\citet{2010PASP..122.1035A} before further reduction.  Bad-pixel
masking, geometric distortion correction, cosmic ray rejection, image
stacking, and resampling were performed on the flat-fielded images
using the MultiDrizzle program \citep{multi}. An optimised resampling
scale of 0.03'', a Gaussian drizzle kernel, and pixfrac = 0.8 were
used to avoid aliasing in the point spread function
\citep{2007ApJS..172..203R}.

Essentially the same field had been observed with ACS already in December
2006 in the F606W filter as part of GO-10881 (PI: Smith), a snapshot
programme targeting clusters at $z{\sim}0.2$. \source\ was erroneously
included in this programme due to confusion with the foreground
cluster A\,3192 ($z{=}0.168$). The relative locations, X-ray
luminosities, and masses of \source\ and A\,3192 are discussed in more
detail in an Appendix to this paper.

\subsubsection{Groundbased spectroscopy}

Likely cluster members as well as potential strong-lensing features in
\source\ were targeted with the Low Resolution Imaging Spectrograph
(LRIS, \citealt{1995PASP..107..375O}; \citealt{2010SPIE.7735E..26R})
on the Keck-I telescope on Mauna Kea in January 2010, November 2011,
and December 2011. With the 600 line/mm grating blazed at 7500 \AA\ on
the red side of the spectrograph, the 6800 \AA\ dichroic as a beam
splitter, and the 300/5000 grism on the blue side, a total of six
multi-object spectroscopy masks targeting 84 unique objects were
observed for exposure times ranging from 3$\times$600 s for galaxies
to 3$\times$1800 s for strong-lensing features. Standard data
reduction procedures including bias subtraction, flat-fielding, sky
subtraction and wavelength calibration were performed before the
extraction of one-dimensional spectra for each object.

\begin{figure*}
    \includegraphics[width=18cm]{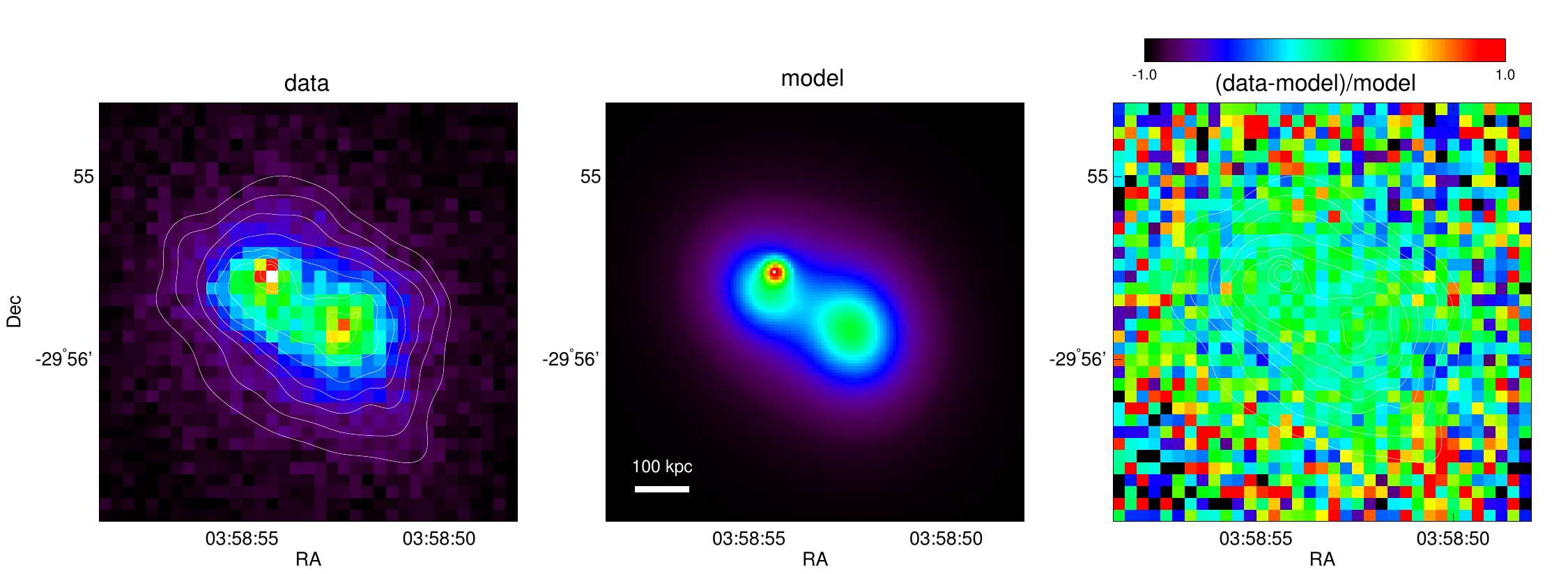}\\[-2mm]
    \caption{Left: X-ray surface brightness distribution. Middle:
      Best-fit spatial model of the surface brightness
      distribution. Right: Residuals.  The model is shown in full
      resolution, while the data and residuals are binned into 4$''$
      pixel. Contours overlaid on the data and residuals show the
      adaptively smoothed X-ray emission. The compact core of the NE
      cluster is clearly visible in the two leftmost panels. Note that
      all the excess over 1.0 in the residuals is shown in
      red.}\mbox{}\\[-5mm]
    \label{fig:spatial}
\end{figure*}

\section{Data analysis and results}
\label{sec:results}

\subsection{X-ray spatial modeling}\label{sec:xsb}

The X-ray surface brightness distribution of
\source\ (Fig.~\ref{fig:overlay}) shows two distinct and well
separated peaks, embedded in a common halo, as well as an extremely
compact core in the north-eastern (NE)
component. 
%which we tentatively identify as a cool core. 
We therefore attempt to parametrize the
observed emission with a three-component model comprised of two
elliptical $\beta$-models for the two sub-clusters and one spherical
2-D Gaussian model for the compact core. A constant, given by the
average value of the exposure-corrected, merged blank-sky image, is
added to account for background emission. The surface brightness
profile $S(\hat{r})$ of each $\beta$-model is described by
\begin{eqnarray}
S(\hat{r})&=&S_0  \left [ \left (1+\frac{\hat{r}}{r_0} \right )^2 \right ]
^{-3\beta+0.5}\\
~~ \hat{r}&=& \sqrt{x^2+ \frac{y^2}{(1-\epsilon)^2}} \nonumber
\label{eq:eq1}
\end{eqnarray}
where x and y are aligned with the major and minor axes of the
ellipse, $\epsilon$ is the ellipticity, $\hat{r}$ is the projected
radius on the sky, $r_0$ is the core radius, $S_0$ is the central
surface brightness, and $\beta$ is the power index.

Acknowledging the small number of X-ray photons detected at larger
cluster-centric radii, we use the Cash statistic during the fit;
unlike for $\chi^2$, the goodness of fit can not be quantified.
Fig. \ref{fig:spatial} shows the data, our model, and the fit
residuals, and demonstrates that our model provides an adequate
description of the observed emission. The best-fit parameters are
given in Table~1.

\begin{table*}
 \caption{Best-fit parameters of the 2-D $\beta$-model for the NE and
   SW sub-clusters, and of the 2-D gaussian model for the compact core
   in the NE sub-cluster}
 \label{tb:1}
 \begin{tabular}{ccccc} 
\hline $r_0$ & $S_0$ & $\epsilon$\tablenotemark{1} & $\theta$ & $\beta$ \\
kpc & photon cm$^{-2}$ s$^{-1}$ arcsec$^{-2}$ & & & \\
 \hline
99.7$^{+5.7}_{-7.5}$ & 4.48$^{+0.25}_{-0.20}$ $\times$10$^{-7}$
 & 0.11$\pm 0.03$ & 126.9$^{\circ}$ $^{+6.8^{\circ}}_{-7.0^{\circ}}$
 & 0.73$^{+0.19}_{-0.20}$ \\
 \hline \\
 \hline
$r_0$& $S_0$ & $\epsilon$ & $\theta$ & $\beta$ \\
kpc & photon cm$^{-2}$ s$^{-1}$ arcsec$^{-2}$ & & & \\
 \hline
128.2$^{+6.9}_{-6.2}$ &5.11$^{+0.16}_{-0.15}$ $\times$10$^{-7}$
& 0.13$\pm 0.02$ & 109.3$^{\circ}$ $^{+5.6^{\circ}}_{-4.2^{\circ}}$
& 0.84$^{+0.20}_{-0.19}$ \\
 \hline \\
 \hline
FWHM &$S_0$ \\
kpc & photon cm$^{-2}$ s$^{-1}$ arcsec$^{-2}$ \\
\hline
35.3$^{+2.9}_{-2.7}$ & 6.25$^{+0.78}_{-0.72}$ $\times$10$^{-7}$ \\
 \hline

\tablenotetext{1}{ellipticity $\epsilon{=}1-b/a$, where a and b are
  the major and minor axis, respectively.}
%\tablenotetext{2}{position angle}

 \end{tabular}
\end{table*}

\subsection{X-ray spectral modeling}\label{sec:spectral}

\begin{figure}
  \begin{center}
    \includegraphics[width=8.5cm]{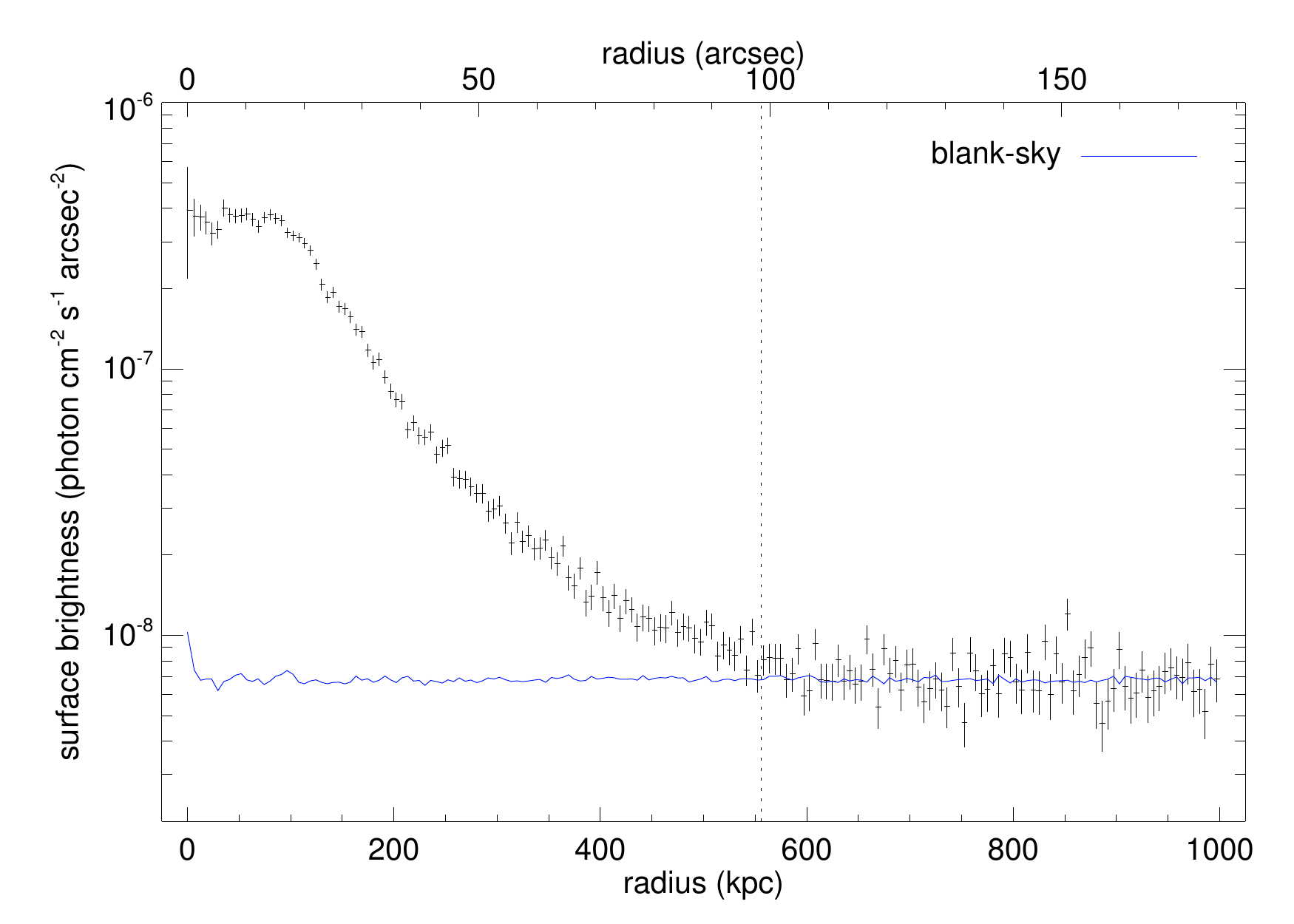}\\[-2mm]
    \caption{Radial profile of the X-ray surface brightness, computed
      relative to the centre of a single-component $\beta$-model. The
      blue line shows the background predicted from appropriately
      scaled blank-sky observations. The vertical dotted line marks
      the radius at which the observed emission becomes statistically
      indistinguishable from the background.}\mbox{}\\[-5mm]
    \label{fig:profile}  
  \end{center}
\end{figure}

In order to determine a suitable region for the extraction of photons
for X-ray spectroscopy, we perform a second spatial fit to the
surface-brightness distribution, this time using a single 2-D
$\beta$-model. A radial surface-brightness profile is then extracted
using the best-fit centre of this model as the centre position, as
shown in Fig. \ref{fig:profile}.

From this profile, we determine the radius at which the
surface-brightness profile becomes statistically indistinguishable
from the blank-sky background. At $r{=}0.556$ Mpc (98.4$''$), the
signal-to-noise ratio (S/N) of the global cluster emission is
maximised, such that extending any measurement to larger radii would
predominantly add background.

\begin{figure}
  \begin{center}
    \includegraphics[width=8.5cm]{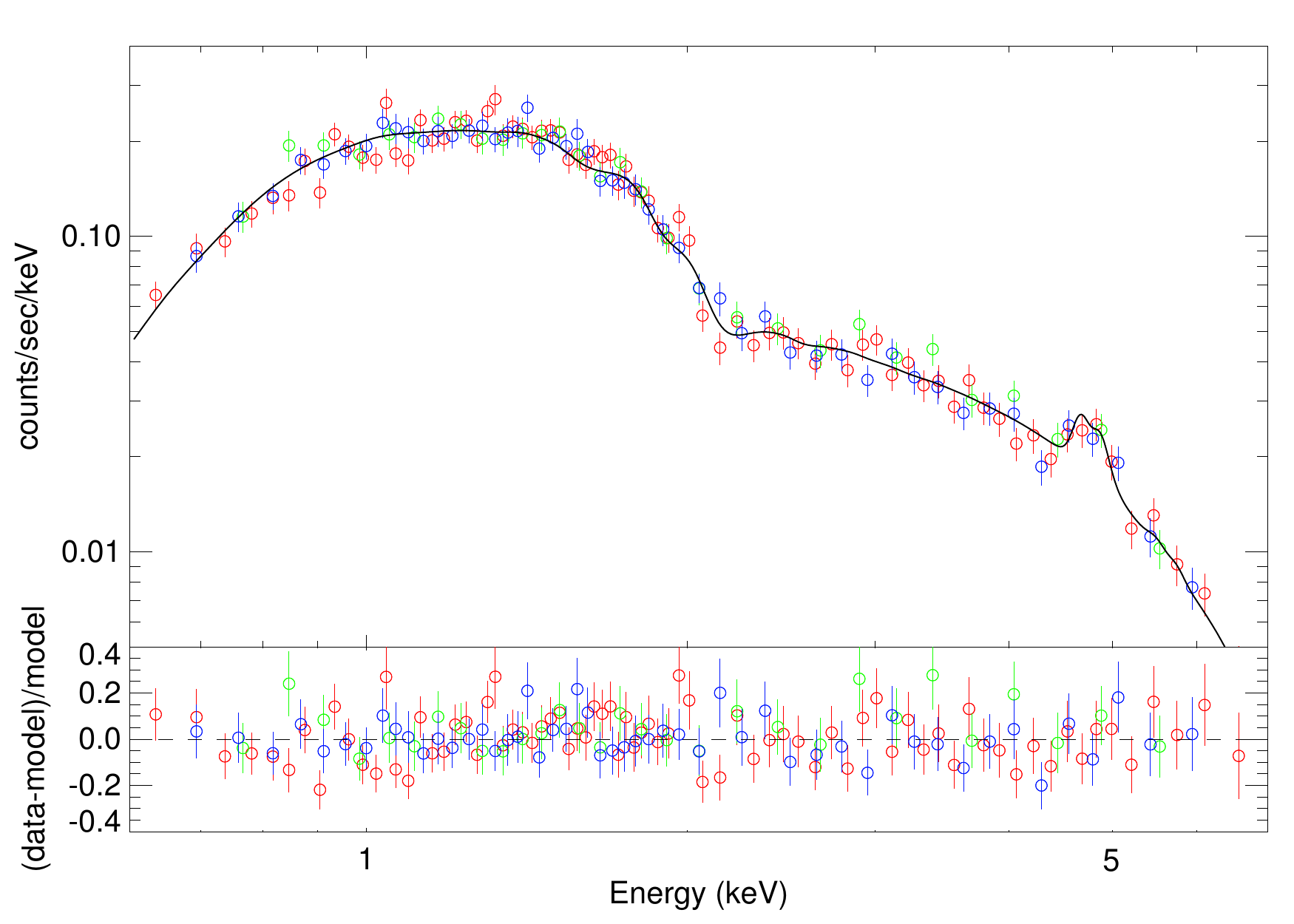}\\[-2mm]
    \caption{X-ray spectrum, best-fit plasma model, and residuals. The
      spectral fitting is actually performed separately but
      simultaneously on the three event files. Here we overplot the
      three spectra, where green, red, and blue data points are from
      ObsID of 11719, 12300, and 13194, respectively.}
    \label{fig:xspectrum}\mbox{}\\[-5mm]
  \end{center}
\end{figure}

Within this radius, we measure a total of 17200 net photons and 1990
background photons in the 0.5--7.0 keV band. All spectral fits are
performed in {\sf Sherpa} \citep{2001SPIE.4477...76F}, using the plasma
model of \citet{1985A&AS...62..197M} combined with the photoelectric
absorption model of \citet{1983ApJ...270..119M}. Fits are performed
simultaneously on the three datasets, using $\chi^2$ statistics. Of
the five fit parameters, we freeze two, namely the cluster redshift
(set to $z=0.428$), and the absorption which we fix at the Galactic
value of $1.00 \times 10^{20}$ cm$^{-2}$
\citep{1990ARA&A..28..215D}. Fitting for the remaining three model
parameters, we find a global temperature of $9.55^{+0.58}_{-0.37}$
keV, a metal abundance of $Z{=}0.27\pm 0.06$, and an unabsorbed X-ray
flux of $(4.05 \pm 0.12)\times 10^{-12}$ erg s$^{-1}$ cm$^{-2}$ in the
quoted energy band. The data, best-fitting model, and residuals are
shown in Fig.~\ref{fig:xspectrum}; the goodness of fit as quantified
by $\chi^2$ per degree of freedom is 0.80.  Thawing the absorption
leads to a best-fit value of $(1.85^{+1.06}_{-1.04}) \times 10^{20}$
cm$^{-2}$ which is fully consistent with the Galactic value; all other
best-fit values are also unchanged within their 1$\sigma$ errors. In
order to explore whether multi-phase gas is present, we also perform a
fit with a two-component plasma model. The resulting best-fit results
are indistinguishable from the ones obtained with a single-phase
model, as the best-fit amplitude of the second component is
negligible. Finally, we also fit for the cluster redshift and measure
$z=0.420^{+0.014}_{-0.020}$ from the redshifted 7 keV Fe line clearly
visible in Fig.~\ref{fig:xspectrum}, in excellent agreement with the
value derived from optical galaxy spectroscopy.

The rest-frame X-ray luminosity of the cluster within the region
selected for X-ray spectroscopy is measured to be $(2.56 \pm 0.06)
\times 10^{45}$ erg s$^{-1}$, $(1.67_{-0.02}^{+0.01})\times 10^{45}$
erg s$^{-1}$, $(1.90 \pm 0.07)\times 10^{45}$ erg s$^{-1}$, and
$(4.04_{-0.13}^{+0.16} )\times 10^{45}$ erg s$^{-1}$ in the 0.5--7.0
keV, 0.1--2.4 keV, 2--10 keV, and bolometric passbands,
respectively. The quoted uncertainties are the nominal 1$\sigma$
errors propagated from the errors of the temperature and normalisation
of the spectral fit. We can extrapolate these luminosities to larger
radii using the best-fit spatial model. From the X-ray
temperature\footnote{We here use the ``core-excised'' temperature of
  $10.01_{-0.65}^{+0.75}$ keV, determined within a radius of 98.4''
  but excluding regions 1, D, and E defined in
  Section~\ref{sec:tmap}.}, we estimate $r_{500}$, $r_{1000}$, and
$r_{2500}$ using the formula in \citet{2002A&A...389....1A},

\begin{eqnarray}\label{eqn:delta}
R_{\Delta} &=& 3.80\beta_{T}^{1/2}\Delta_{z}^{-1/2}(1+z)^{-3/2} \\
& & \times (kT/10~\mathrm{keV})^{1/2}h_{50}^{-1} ~~\mathrm{Mpc} \nonumber 
\end{eqnarray}
Here $\beta$ = 1.05 is the normalisation of the virial relation, i.e.,
$GM/(2R_{vir})=\beta_{T}kT$, and $\Delta_z =
(\Delta\Omega_{m_0})/(18\pi^2\Omega_m(z))$ with $\Delta$ being the
desired overdensity. Setting $\Delta$ to 500, 1000, and 2500 yields
$r_{500}{=}1.35$ Mpc (239$''$), $r_{1000}{=}0.95$ Mpc (169$''$), and
$r_{2500}{=}0.60$ Mpc (107$''$). Our measurement aperture of
$r{=}0.56$ Mpc (98$''$) corresponds to $\Delta \sim 2900$.

Using the best model from our two-dimensional fit to the spatial data
we thus find $L_X(<r_{500})$ of $2.69 \times 10^{45}$ erg s$^{-1}$,
$1.76 \times 10^{45}$ erg s$^{-1}$, $2.00 \times 10^{45}$ erg
s$^{-1}$, and $4.24 \times 10^{45}$ erg s$^{-1}$ for the for 0.5--7.0
keV, 0.1--2.4 keV, 2--10 keV and bolometric passbands, respectively.

\subsection{Gas mass and total mass}

The normalisation factor of the spectral fit and the best-fit spatial
model allow us to estimate the total X-ray gas mass within a given
region. The normalisation of the {\sf mekal} model is given by
\begin{equation}                                                               
{\tt norm}={10^{-14}\over4 \pi {D_A}^2(1+z)^2}\int n_e n_H dV
\label{eq:norm}
\end{equation}  
where the emission measure, $\int n_e n_H dV$,  can be written as
\begin{equation}
{\tt EM} =  (n_H/n_e) \int n_e(\vec{r})^2 dV,
\end{equation}
and the gas mass is given by
\begin{equation}
M_{gas} = \mu' m_p (n_H/n_e) \int n_e(\vec{r})^2 dV.
\end{equation}
Here $m_p$ is the proton mass, $\mu'$ = 1.347, and $n_H/n_e$ = 
0.852 for an ionized plasma with a metallicity of 0.3 of the solar 
value \citep{2002A&A...389....1A}.

\begin{figure}
  \begin{center}
    \includegraphics[width=8.5cm]{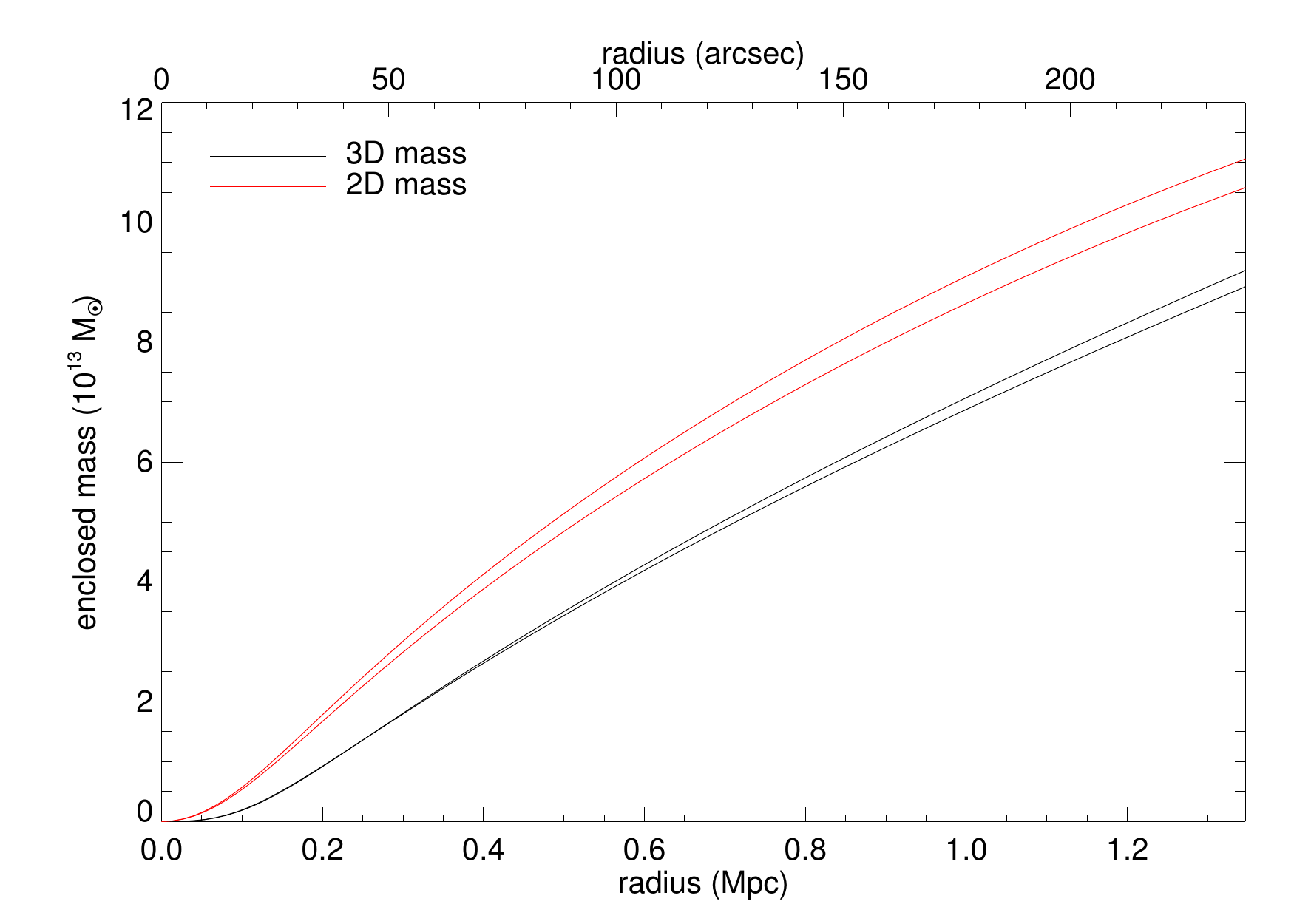}
    \caption{Profiles of 3D and 2D gas mass derived for oblate and
      prolate geometries. For each set of curves, the upper one
      corresponds to oblate geometry. The vertical dotted line marks
      the radius from Fig.~3 at which our existing observations fail
      to detect significant emission from the cluster.}
    \label{fig:mass}
  \end{center}
\end{figure}

We deproject the surface-brightness distribution determined by the
spatial model in Section \ref{sec:xsb} to a three-dimensional gas
density profile
\begin{eqnarray}\label{eqn:beta3d}
\rho(r) &=& \rho_0 \left [ \left (1+\frac{r}{r_0} \right )^2 \right ]
^{-\frac{3}{2}\beta}\\ 
~ r &=& \sqrt{x^2+ \frac{y^2}{(1-\epsilon_y)^2}+
  \frac{z^2}{(1-\epsilon_z)^2}} \nonumber
\end{eqnarray}
where x and y are aligned with the major and minor axes of the ellipse
projected on the plane of the sky, z is along our line of sight, and
$\epsilon_y$, $\epsilon_z$ are the corresponding ellipticities. We can
then numerically integrate the right-hand side of Eqn.~\ref{eq:norm}
to obtain the central density $\rho_0$ (and $n_{e0}$) of each of the
three spatial components. For the deprojection, however, we need to
assume a geometry for the gas distribution along the line of sight. We
assume that the two elliptical gaseous halos of our best-fit 2D
spatial model are projections of 3D spheroids which have two of their
three axes of symmetry lying in the plane of the sky.
Fig. \ref{fig:mass} shows the resulting gas mass for the two extreme
cases of oblate and prolate geometries (corresponding to
$\epsilon_z{=}0$ and $\epsilon_z{=}\epsilon_y$, respectively) as a
function of radius\footnote{We adopt a density cutoff at $r {=}
  r_{200}$, the approximate virial radius.}. The total 3D gas mass
derived within $r_{500}$ is $(9.20^{+1.44}_{-1.22}) \times 10^{13}
M_{\odot}$ for oblate and $(8.93^{+1.42}_{-1.19}) \times 10^{13}
M_{\odot}$ for prolate geometries, respectively. For the 2D masses,
these values increase to $(1.11^{+0.18}_{-0.15}) \times 10^{14}
M_{\odot}$ and $(1.06^{+0.18}_{-0.15}) \times 10^{14} M_{\odot}$. The
quoted uncertainties are the nominal 1$\sigma$ errors propagated from
the spatial and spectral fit results. Combining the mass ranges for
the two geometries, we find $M^{\rm 3D}_{\rm gas}(<r_{500}) = (9.18
\pm 1.45) \times 10^{13} M_{\odot}$.

An estimate of the total gravitational mass can be derived from the
gas mass by assuming a value for the gas mass fraction. The latter
depends, however, on the radius from the cluster centre as well as on
the total cluster mass (or via proxy its gas temperature), and is also
known to vary considerably between clusters. Average values of $f_{\rm
  gas}(<r_{500}){\sim}0.05$ are found for groups of galaxies
\citep{2012NJPh...14d5004S}, rising to 0.13 for the most massive
clusters \citep{2009ApJ...692.1033V}, with a scatter of typically
$\pm0.02$ for individual clusters of comparable mass
\citep{2012NJPh...14d5004S}. A detailed analysis of a disturbed,
optically selected cluster of intermediate mass by
\citet{2012ApJ...748..120D}, for instance, yields relatively low
values of $f_{\rm gas}(<r_{2500}){=}0.06$ and $f_{\rm
  gas}(<r_{500}){=}0.1$, respectively, possibly due to the ongoing
merger.

We here use an estimate of $f_{\rm gas}({<}r_{500}){=}0.082$ for
\source\ to convert gas mass to total gravitational mass, a value that
is derived from a comparison of the radial profile of the gas mass
with that of the total mass derived from our strong-lensing analysis
of the system (see Section~\ref{sec:lens_model}).  We thus find
$M^{\rm 3D}_{tot}(<r_{500}) = (1.12 \pm 0.18) \times 10^{15}
M_{\odot}$. The corresponding 2D masses are $M^{\rm 2D}_{\rm
  gas}(<r_{500}) = (1.10 \pm 0.19) \times 10^{14} M_{\odot}$, and
$M^{\rm 2D}_{tot}(<r_{500}) = (1.34 \pm 0.23) \times 10^{15}
M_{\odot}$. All errors include only statistical and modeling
uncertainties.

\subsection{X-ray temperature map}\label{sec:tmap}

\begin{figure*}
  \begin{center}
    \includegraphics[width=15.5cm]{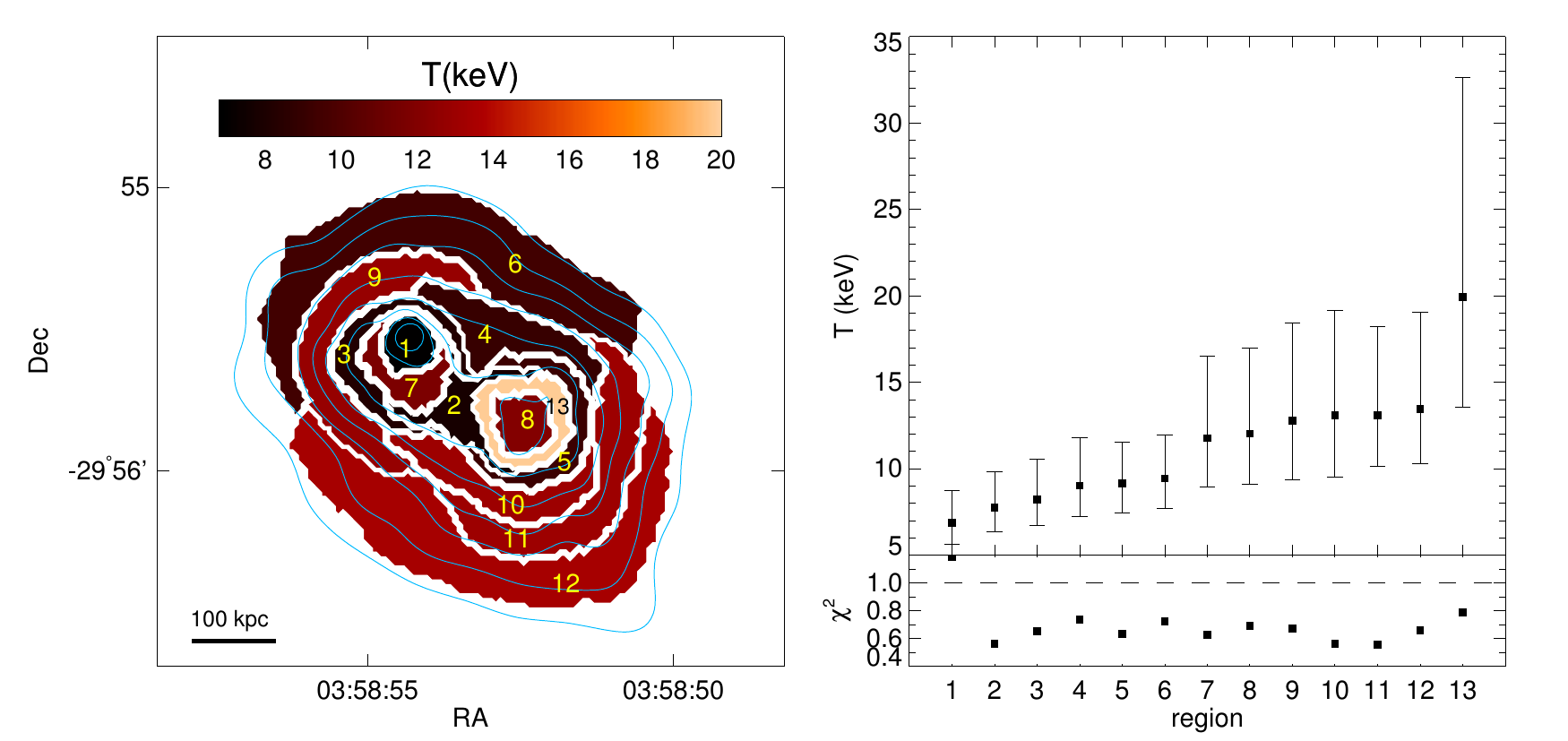}
    \includegraphics[width=15.5cm]{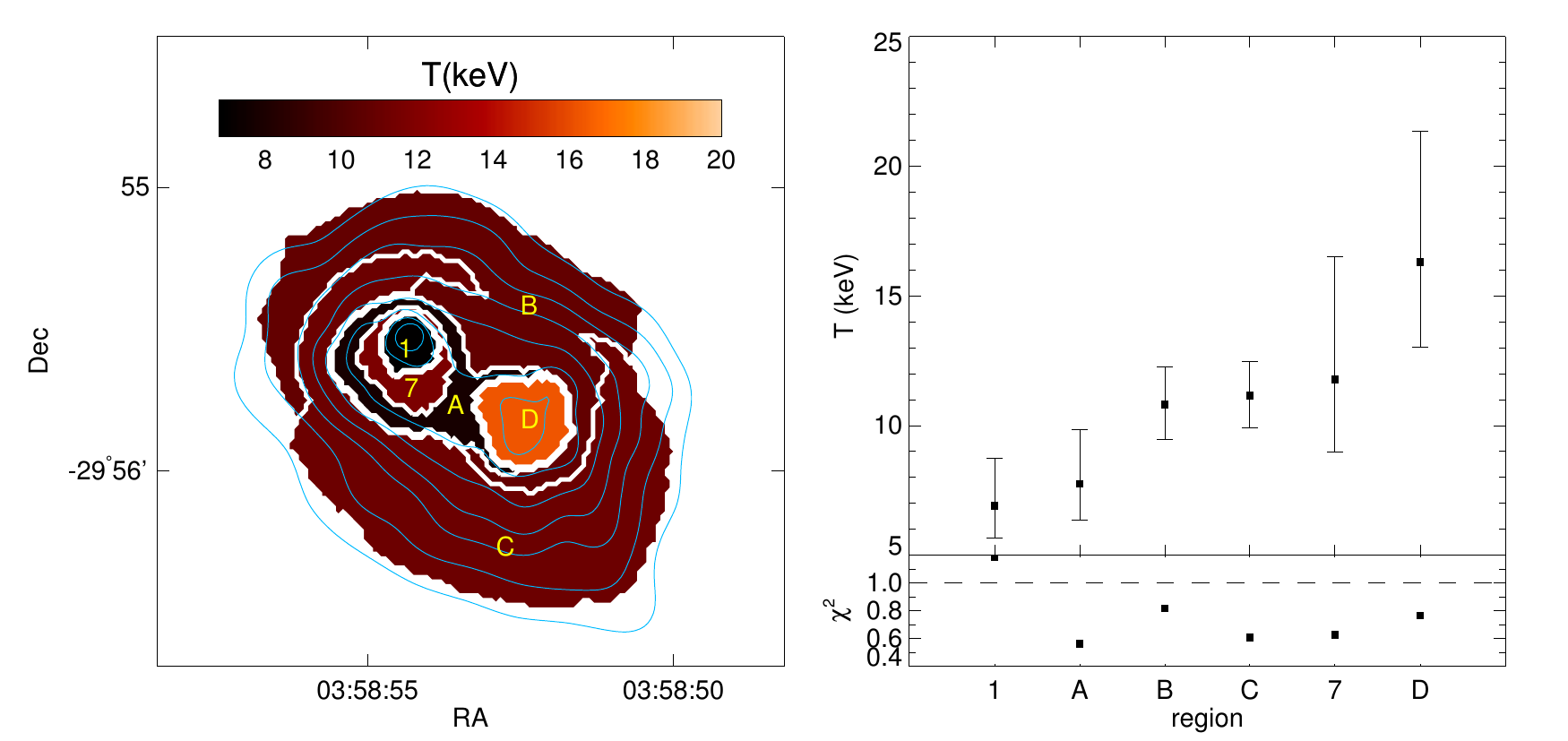}
    \includegraphics[width=15.5cm]{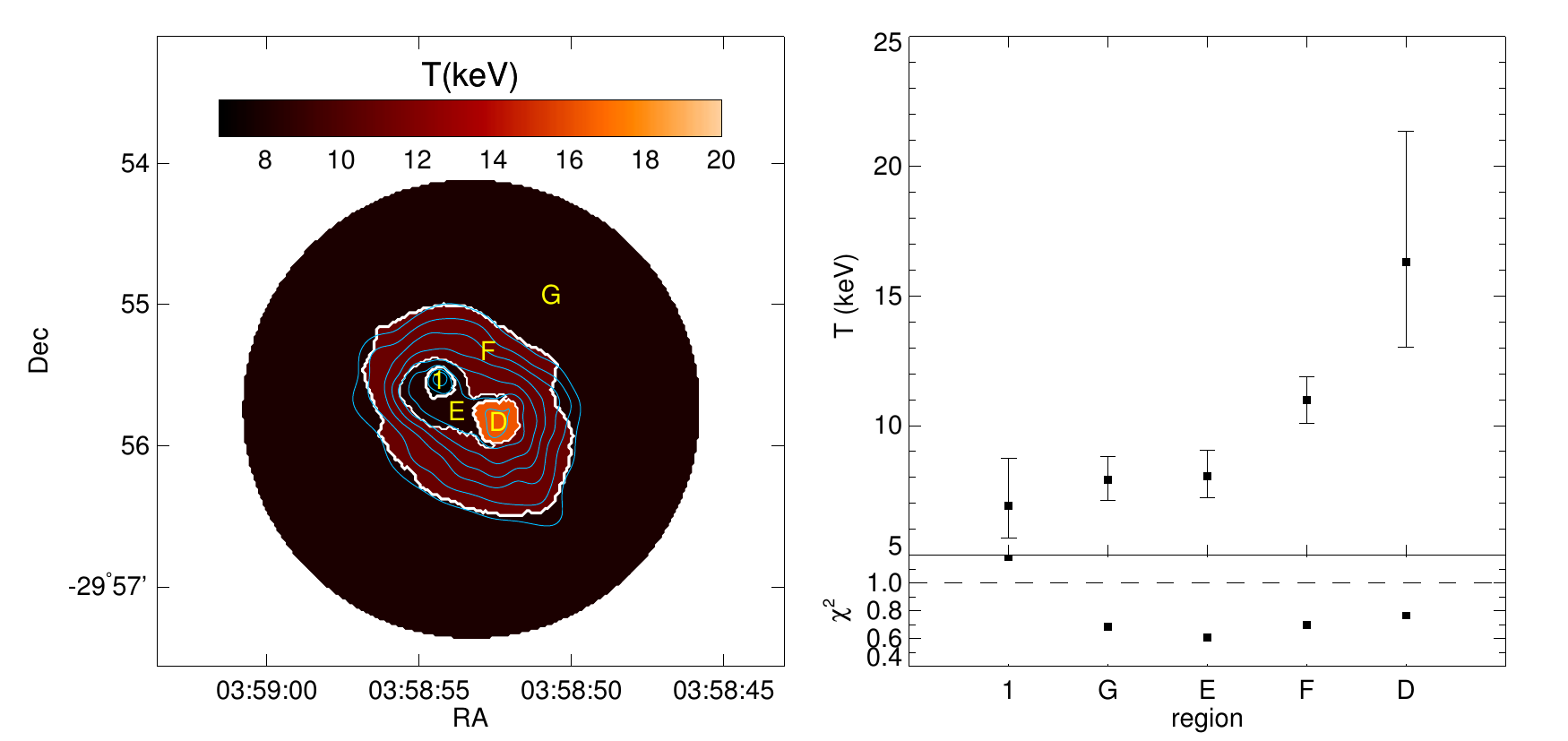}
    \caption{Left panels: gas temperatures in different regions
      estimated by fitting plasma and photoelectric absorption models
      to the respective X-ray spectra. Temperatures are colour coded
      as indicated by the shown colour bar. Purple contours show the
      adaptively smoothed X-ray emission. Right panels: the best-fit
      temperatures and reduced $\chi^2$ values for each region; labels
      on the x axis correspond to the naming of regions as shown in
      the left panel. The three sets of panels, from top to bottom,
      show the results of our first, second, and third iteration of
      the temperature map.  In the bottom panels, we include a region
      outside of the area shown in the previous two maps; we use this
      region to measure the ambient gas temperature at large
      radii. Merged regions as labelled are defined as follows: A =
      2\&3, B = 4\&5\&6, C = 9\&10\&11\&12, D = 8\&13, E = 7\&A, F =
      B\&C. }
    \label{fig:map}
  \end{center}
\end{figure*}

In order to explore temperature variations across the emission region,
we first define sub-regions using the ``contour binning'' algorithm
(\emph{contbin}, \citealt{2006MNRAS.371..829S}) for a given S/N, and
then perform spectral fits for each region. These fits follow the
prescription provided in Section~\ref{sec:spectral} except that the
metal abundance is fixed at a value of 0.3, a typical value of
clusters, leaving only the gas temperature and the normalisation as
free parameters. Requiring a S/N of at least 30, corresponding to
typically at least 940 net photons in each region, yields the
temperature map shown in the upper left panel of Fig.~\ref{fig:map}.
We then follow the approach described in \citet{2009ApJ...693L..56M}
and merge adjacent regions that exhibit similar temperatures.  A new
set of spectral fits is performed on these larger regions with
improved photon statistics to obtain new maps with lower resolution
but higher fidelity.

The middle panels of Fig.~\ref{fig:map} show the results of the third
iteration in which adjacent regions of similar temperature have been
merged twice. Region D in the middle panel deserves a special note.
It results from the merging of regions 8 and 13 in the top panels
which were combined because their temperatures are in fact consistent
within the large errors. However, the large uncertainty of the
temperature measured for the merged region strongly suggests the
presence of multi-phase gas. A large uncertainty in the best-fit
temperature is also seen for region 7 (both top and middle panels)
which is not merged with any other adjacent region. The photon
statistics in both region 7 and region D are, however, insufficient to
constrain a two-temperature model.

The bottom panels of Fig.~\ref{fig:map}, finally, show the temperature
distribution obtained when this procedure of merging adjacent regions
of similar temperature is applied further.  Here we combine the two
ambient regions labeled B and C, as well as the regions labeled 7 and
A. In addition, we also show in these panels a region inside our
large-scale spectral aperture (see Section~\ref{sec:spectral}), but
outside the regions labelled in the remainder of this figure. The
right-hand panels show that the differences between the best-fit gas
temperatures of the NE sub-cluster (1), the SW sub-cluster (D), and
the ambient region (F) are statistically significant.

\begin{figure*}
    \includegraphics[width=7cm]{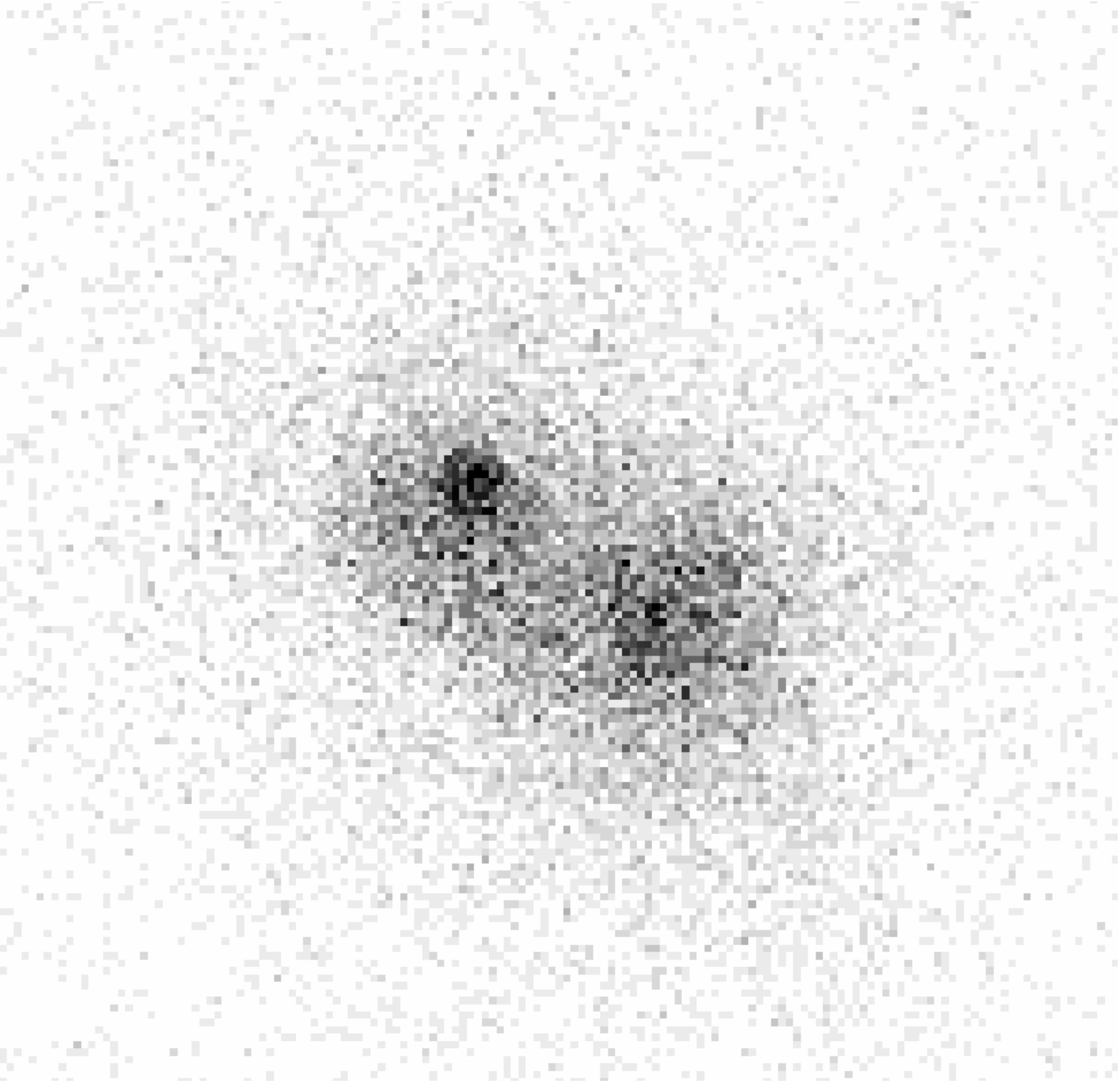}\hspace*{2cm}
    \includegraphics[width=7cm]{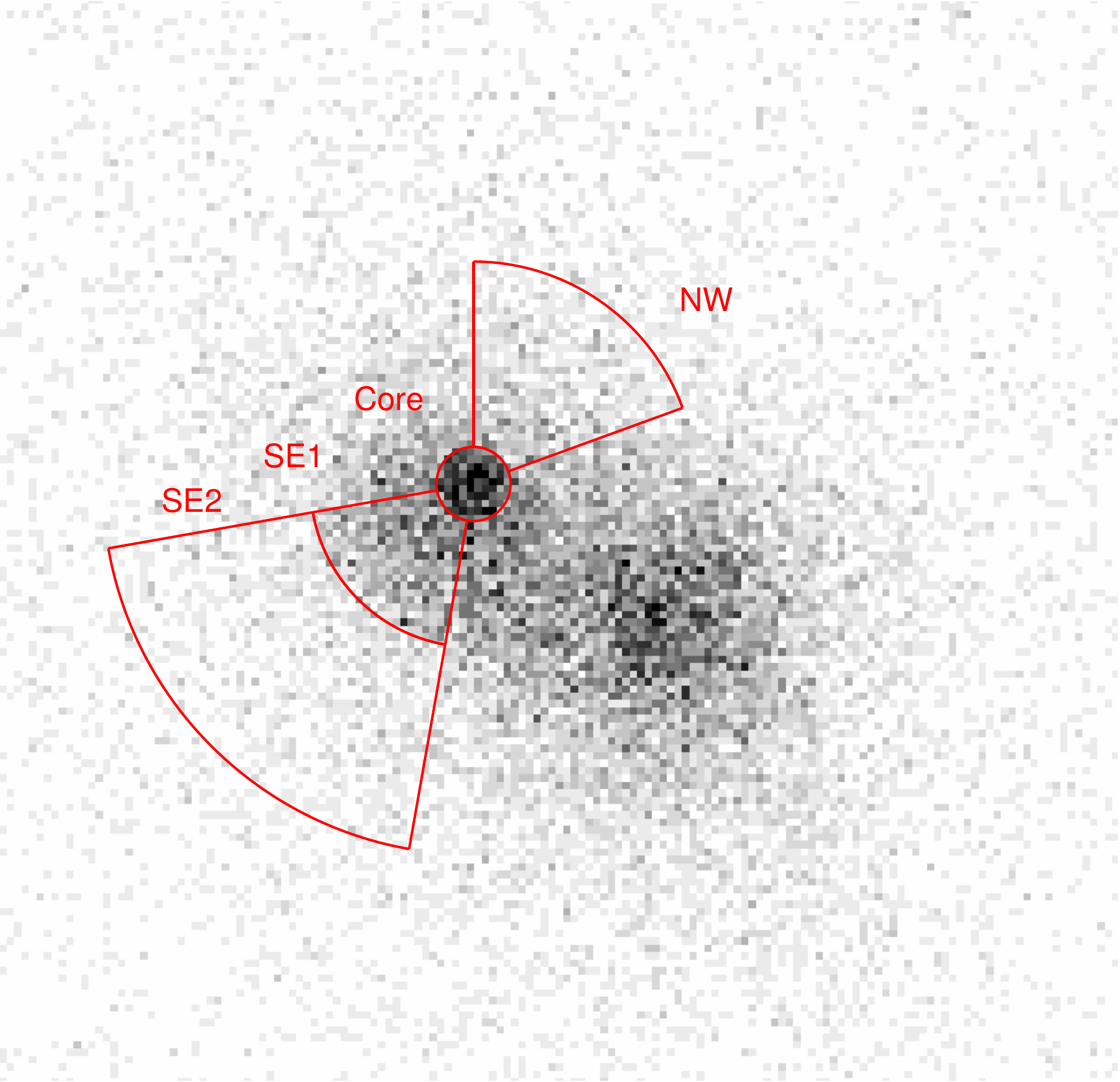}
    \caption{Left: exposure-corrected image of \source\ in the 0.5-7.0
      keV energy band. Right: as on left, but with regions used to
      analyse the surface brightness edges overlaid.}
    \label{fig:sectors}
\end{figure*}

\begin{figure}
  \begin{center}
    \hspace*{-5mm}\includegraphics[width=9.5cm]{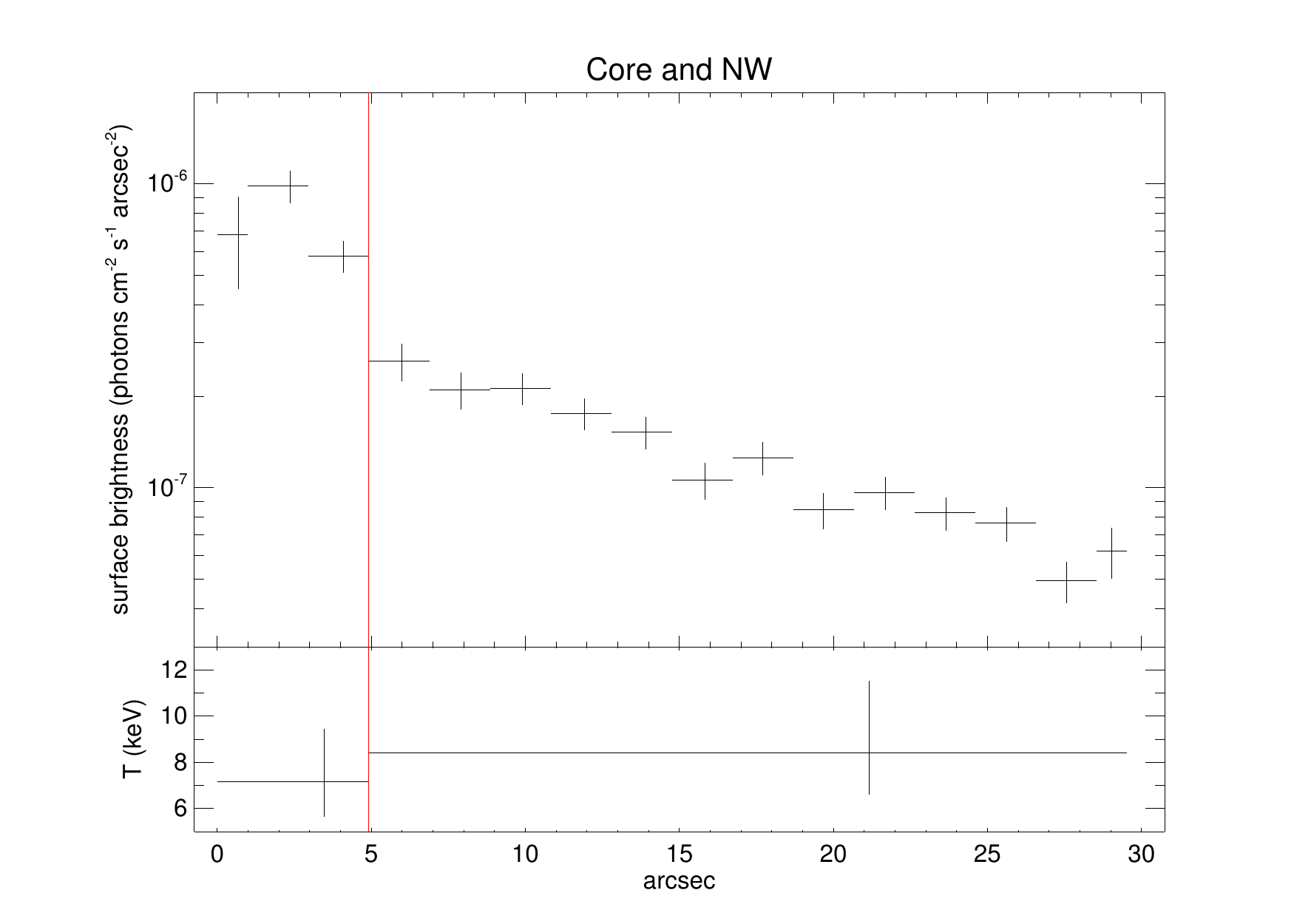} \mbox{}\\[-9mm]
    \hspace*{-5mm}\includegraphics[width=9.5cm]{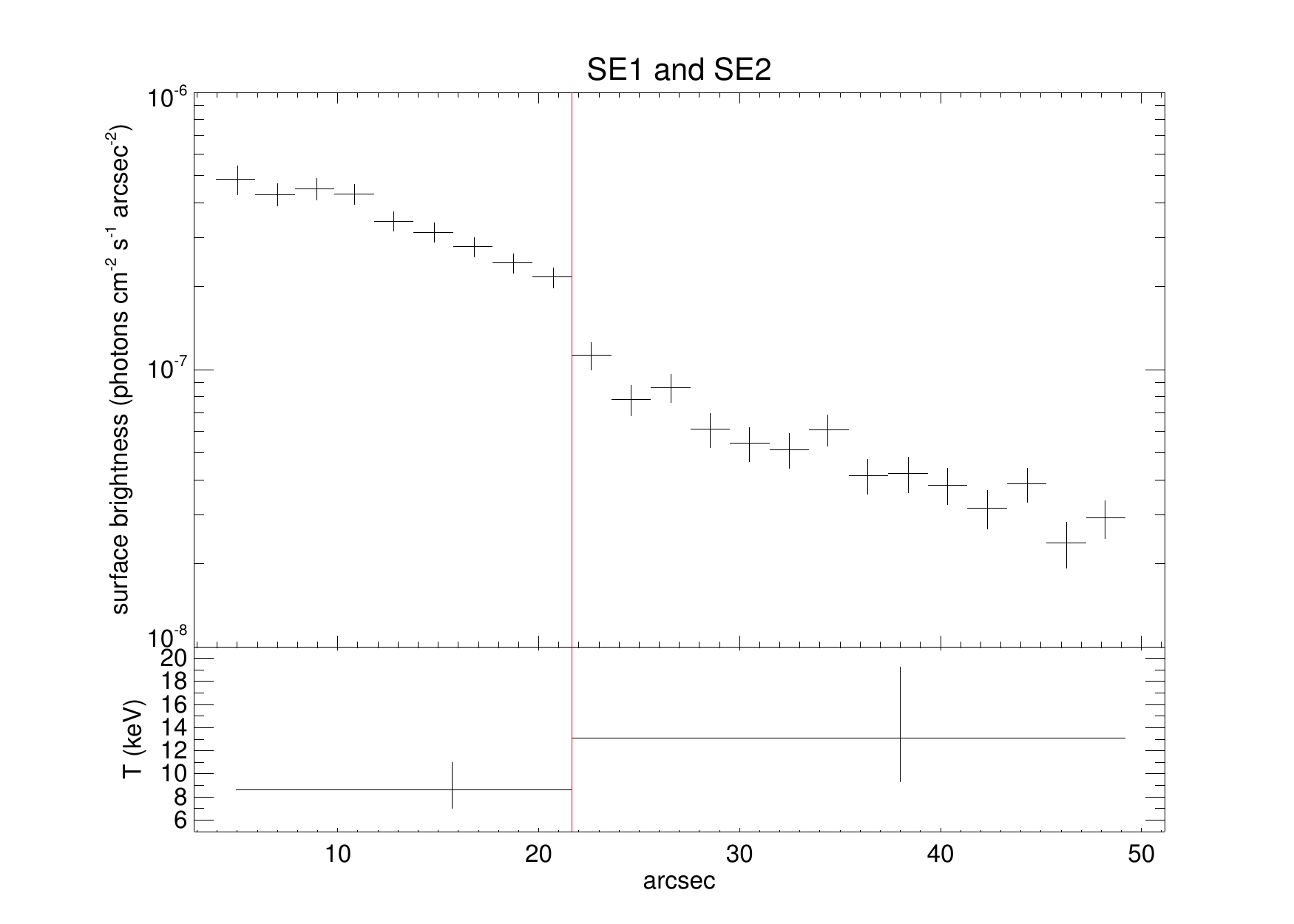}
    \caption{The surface brightness (in the energy range 0.5-7.0 keV)
      and temperature profiles for the two edges extracted from the
      four regions defined in Fig. \ref{fig:sectors}.}
    \label{fig:front}
  \end{center}
\end{figure}

\subsection{Putative cold fronts}\label{sec:coldfront}

Visual inspection of our X-ray image of \source\ suggests the presence
of two surface-brightness discontinuities in the NE region of the
cluster. One is at the north-western side of the compact core in the
NE cluster; the other one is at the south-eastern side of the same
sub-cluster. The clear excess of the X-ray emission in this core
region over the $\beta$-model describing the NE cluster on larger
scales (Section~\ref{sec:xsb} and Fig.~\ref{fig:spatial}) in
conjunction with the low ICM temperature of about 7 keV measured there
(Fig.~\ref{fig:map}) identify this feature as a likely cool core. To
allow our readers to convince themselves of the plausibility of the
proposed discontinuities on either side of the potential cool core, we
show in Fig.~\ref{fig:sectors}, side by side, the raw data and the raw
data with four regions (Core, NW, SE1 and SE2) marked that we define
to measure the temperature changes across the surface brightness
edges.

The measured surface-brightness and temperature distributions across
the two putative fronts are shown in Fig.~\ref{fig:front}. We find a
surface-brightness discontinuity that is significant at the 3.0 and
3.3$\sigma$ confidence level, respectively. The photon statistics are,
however, not good enough to tightly constrain the gas temperature
locally on either side of these discontinuities. Given that, for both
fronts, the best-fit temperature is slightly lower on the denser side,
consistent with pressure equilibrium across the feature, we propose
that these two fronts are cold fronts.

\begin{figure}
\begin{center}
\includegraphics[width=80mm]{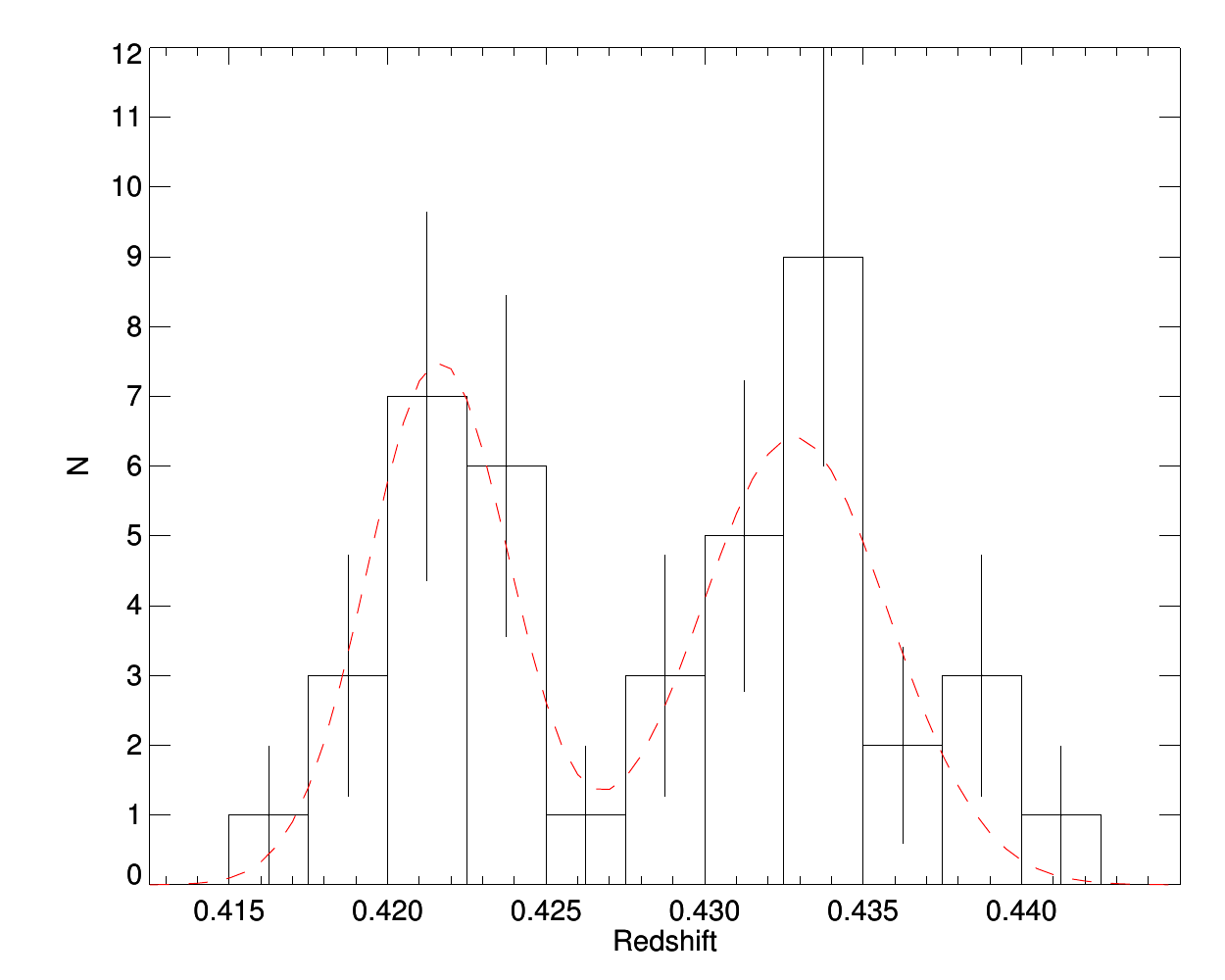}
\caption{Distribution of spectroscopic redshifts of the cluster
  galaxies in \source\ as determined from our LRIS observations. The
  shown error bars assume Poisson statistics.
  The dashed line
  describes a double-Gaussian distribution with velocity dispersions
  of $670\rm~km~s^{-1}$ (predominantly describing the SW subcluster)
  and $890\rm~km~s^{-1}$ (predominantly the NE subcluster) centred on
  the redshifts of $z=0.4216$ and $z=0.4328$. }
\label{fig:z_hist}
\end{center}
\end{figure}

\begin{table*}
\caption{Spectroscopic galaxy redshifts in the field of \source\ except the members of A\,3192.}\label{lris_redshift}
\begin{tabular}{@{\hspace{2mm}}c@{\hspace{2mm}}c@{\hspace{2mm}}r@{\hspace{3mm}}c@{\hspace{2mm}}c@{\hspace{2mm}}r@{\hspace{4mm}}c@{\hspace{10mm}}c@{\hspace{2mm}}c@{\hspace{2mm}}r@{\hspace{3mm}}c@{\hspace{2mm}}c@{\hspace{2mm}}r@{\hspace{4mm}}c}
\hline
\hline
\multicolumn{3}{c}{R.A.}&\multicolumn{3}{c}{Dec.} & {\it z} & \multicolumn{3}{c}{R.A.} & \multicolumn{3}{c}{Dec.} & {\it z}\\

\hline
03 & 58 & 45.424 & -29 & 55 & 24.03 & 0.3375   &03 & 58 & 53.783 & -29 & 55 & 20.08 & 0.5113 \\
03 & 58 & 46.140 & -29 & 55 & 32.77 & 0.4245   &03 & 58 & 53.783 & -29 & 54 & 56.30 & 0.4201 \\
03 & 58 & 47.693 & -29 & 57 & 8.90  & 0.4362   &03 & 58 & 53.966 & -29 & 55 & 55.64 & 0.4749 \\
03 & 58 & 47.886 & -29 & 56 & 29.94 & 0.4210   &03 & 58 & 54.029 & -29 & 54 & 49.50 & 0.5150 \\
03 & 58 & 49.004 & -29 & 56 & 7.22  & 0.4273   &03 & 58 & 54.127 & -29 & 55 & 31.41 & 0.4340\tablenotemark{a}\\
03 & 58 & 49.685 & -29 & 55 & 59.58 & 0.4423   &03 & 58 & 54.206 & -29 & 54 & 51.43 & 0.2380 \\
03 & 58 & 49.890 & -29 & 26 & 43.43 & 0.2178   &03 & 58 & 54.630 & -29 & 57 & 57.17 & 0.3948 \\
03 & 58 & 50.541 & -29 & 56 & 51.34 & 0.4209   &03 & 58 & 54.856 & -29 & 55 & 55.37 & 1.262 \tablenotemark{b}\\
03 & 58 & 50.673 & -29 & 56 & 34.12 & 0.4220   &03 & 58 & 52.400 & -29 & 55 & 23.77 & multiple\tablenotemark{b} \\
03 & 58 & 50.774 & -29 & 56 & 29.92 & 0.4207   &03 & 58 & 53.553 & -29 & 55 & 38.36 & multiple\tablenotemark{b} \\
03 & 58 & 50.902 & -29 & 55 & 34.28 & 0.4288   &03 & 58 & 55.172 & -29 & 56 & 35.38 & 0.4308 \\
03 & 58 & 50.991 & -29 & 55 & 42.03 & 0.2171   &03 & 58 & 55.589 & -29 & 55 & 39.67 & 0.4346 \\
03 & 58 & 51.350 & -29 & 56 & 20.18 & 1.537\tablenotemark{d} &03 & 58 & 55.995 & -29 & 55 & 11.71 & 0.4400 \\
03 & 58 & 51.744 & -29 & 57 & 16.82 & 0.4287   &03 & 58 & 56.092 & -29 & 57 & 26.17 & 0.4329 \\
03 & 58 & 51.751 & -29 & 54 & 19.27 & 0.4241   &03 & 58 & 56.630 & -29 & 55 & 21.08 & 0.2704 \\
03 & 58 & 51.814 & -29 & 57 & 35.58 & 0.4331   &03 & 58 & 56.681 & -29 & 55 & 30.63 & 0.4384 \\
03 & 58 & 51.827 & -29 & 53 & 28.99 & 0.3380   &03 & 58 & 56.904 & -29 & 54 & 56.83 & 0.4237 \\
03 & 58 & 52.017 & -29 & 56 & 15.04 & 0.4230   &03 & 58 & 56.947 & -29 & 55 & 31.43 & 0.4332 \\
03 & 58 & 52.138 & -29 & 54 & 21.98 & 0.3383   &03 & 58 & 57.024 & -29 & 55 & 12.73 & 0.4354 \\
03 & 58 & 52.170 & -29 & 56 & 4.27  & 0.4233   &03 & 58 & 57.127 & -29 & 55 & 18.88 & 0.3947 \\
03 & 58 & 52.316 & -29 & 56 & 19.04 & 0.4210   &03 & 58 & 57.158 & -29 & 54 & 47.47 & 0.4382 \\
03 & 58 & 52.359 & -29 & 56 & 58.60 & 0.4312   &03 & 58 & 57.181 & -29 & 54 &  2.58 & 0.4176 \\
03 & 58 & 52.575 & -29 & 54 & 56.63 & 0.4341   &03 & 58 & 57.298 & -29 & 55 & 10.10 & 0.4319 \\
03 & 58 & 52.687 & -29 & 56 & 11.13 & 3.07\tablenotemark{c}      &03 & 58 & 57.479 & -29 & 54 & 12.71 & 0.4338 \\
03 & 58 & 52.959 & -29 & 56 & 11.56 & multiple\tablenotemark{c}  &03 & 58 & 57.975 & -29 & 54 & 30.55 & 0.4342 \\
03 & 58 & 50.097 & -29 & 55 & 22.94 & multiple\tablenotemark{c}  &03 & 58 & 58.179 & -29 & 53 & 49.26 & 0.1886 \\
03 & 58 & 52.691 & -29 & 55 & 50.11 & 0.4186   &03 & 58 & 58.366 & -29 & 55 & 38.04 & 0.4233 \\
03 & 58 & 52.785 & -29 & 55 & 27.98 & 0.4172   &03 & 58 & 58.539 & -29 & 54 & 48.05 & 0.4198 \\
03 & 58 & 52.977 & -29 & 55 & 41.94 & 0.4292   &03 & 58 & 58.538 & -29 & 54 & 53.66 & 0.4342 \\
03 & 58 & 52.989 & -29 & 55 & 57.62 & 0.4306   &03 & 58 & 58.887 & -29 & 54 &  7.84 & 0.5542 \\
03 & 58 & 53.605 & -29 & 57 & 4.52  & 0.4204   &03 & 58 & 59.447 & -29 & 54 & 46.63 & 0.4616 \\
03 & 58 & 53.780 & -29 & 54 & 51.59 & 0.5127   &03 & 58 & 59.726 & -29 & 56 & 21.95 & 0.4320 \\
\hline
\hline
\tablenotetext{a}{The north-eastern BCG} 
\tablenotetext{b}{The triple system A}  
\tablenotetext{c}{The triple system B}
\tablenotetext{d}{The gravitational arc D}

\end{tabular}
\end{table*}

\begin{table*}
\caption{Spectroscopic redshifts of the members of A\,3192 in the field of \source.}\label{lris_redshift2}
\begin{tabular}{@{\hspace{2mm}}c@{\hspace{2mm}}c@{\hspace{2mm}}r@{\hspace{3mm}}c@{\hspace{2mm}}c@{\hspace{2mm}}r@{\hspace{4mm}}c@{\hspace{10mm}}c@{\hspace{2mm}}c@{\hspace{2mm}}r@{\hspace{3mm}}c@{\hspace{2mm}}c@{\hspace{2mm}}r@{\hspace{4mm}}c}

\hline 
\hline

\multicolumn{3}{c}{R.A.}&\multicolumn{3}{c}{Dec.} & {\it z} & \multicolumn{3}{c}{R.A.} & \multicolumn{3}{c}{Dec.} & {\it z}\\

\hline

03 & 58 & 44.513 & -29 & 55 & 46.02 & 0.1750  & 03 & 58 & 51.463 & -29 & 57 & 5.51  & 0.1670 \\
03 & 58 & 46.572 & -29 & 57 & 2.05  & 0.1724  & 03 & 58 & 51.819 & -29 & 53 & 46.38 & 0.1657 \\
03 & 58 & 47.183 & -29 & 56 & 59.43 & 0.1690  & 03 & 58 & 53.788 & -29 & 57 & 46.49 & 0.1666 \\
03 & 58 & 50.901 & -29 & 56 & 52.17 & 0.1634  & 03 & 58 & 57.644 & -29 & 54 &  4.27 & 0.1610 \\
03 & 58 & 51.165 & -29 & 54 & 13.56 & 0.1730  & 03 & 58 & 58.092 & -29 & 54 & 26.61 & 0.1697 \\

\hline
\hline

\end{tabular}
\end{table*}

\begin{figure*}
  \begin{center}
    \includegraphics[width=17cm]{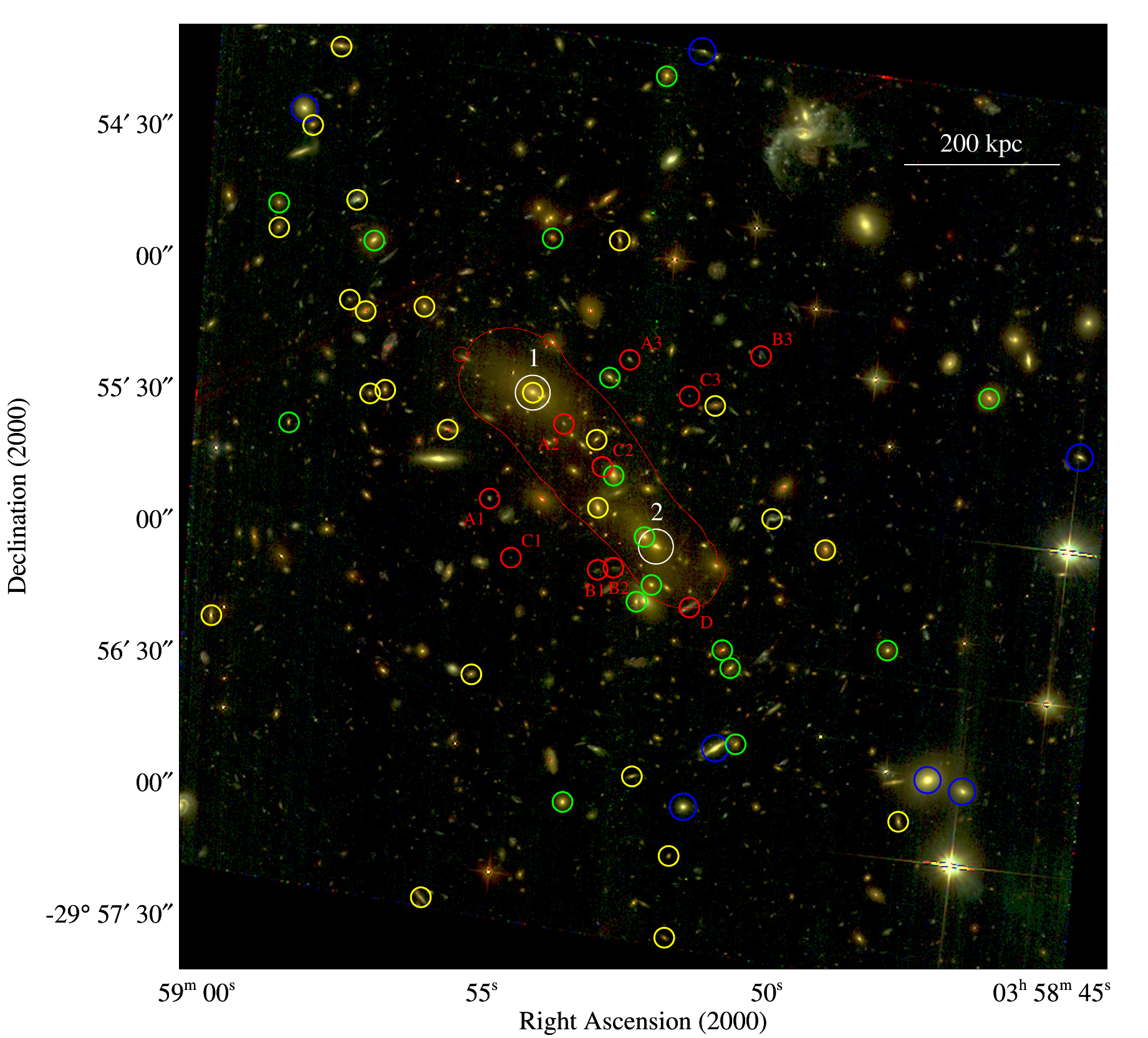}
    \caption{HST colour image (blue: F435W, green:F606W, red: F814W)
      of \source. Green circles are cluster galaxies with redshifts
      below 0.4268. Yellow circles are cluster galaxies with redshifts
      above 0.4268. Red circles highlight the strong-lensing features
      shown in Fig.~\ref{fig:multiples}. Blue circles mark galaxies
      around $z = 0.168$, i.e., members of the foreground system
      A\,3192. White circles mark the locations of the two light peaks
      used for the strong-lensing mass model; the critical line at
      $z=1.262$ for System A (see Section~\ref{sec:lens_model}) is
      shown in red.}
    \label{fig:overlay2}
  \end{center}
\end{figure*}

\subsection{Spectroscopic redshifts}
\label{sec:redshifts}

Spectroscopic redshifts of galaxies are determined by
cross-correlation between reduced 1-D spectra and spectral templates
provided in the IDL routine {\sf SpecPro} developed by
\citet{2011PASP..123..638M}. The exception are three lensed background
galaxies and two emission-line galaxies for which we determine
redshifts by identifying emission lines. Based on the measured
redshifts we identify 41 cluster members with redshifts between 0.4172
and 0.4423, 20 foreground objects, and 9 background objects, three of
which are the gravitational arc and multiple systems discussed in more
detail in Section~\ref{sec:lens_model}. Ten of the foreground galaxies
have redshifts that identify them as members of the foreground cluster
A\,3192 at $z=0.168$. The redshifts of these ten galaxies in A\,3192
are listed in Table~\ref{lris_redshift2}; all others are tabulated in
Table~\ref{lris_redshift}.

We calculate a mean cluster redshift of $0.4284$ and a large radial
velocity dispersion of $1440^{+130}_{-110}\rm~km~s^{-1}$. A histogram
of the cluster galaxies is shown in Fig.~\ref{fig:z_hist}. The
distribution is clearly bimodal and well described by a
double-Gaussian with velocity dispersions\footnote{In keeping with
  results presented in the literature all velocity dispersions quoted
  in this Section lack relativistic corrections. The latter become,
  however, highly significant already at $z\sim 0.2$. Applying the
  correct relativistic conversion between redshift and radial velocity
  leads to greatly reduced values of 892 km s$^{-1}$ for the overall
  velocity dispersion, and 295 and 383 km s$^{-1}$ for the two
  components, respectively.} of ($670\pm 170$)~km~s$^{-1}$ and ($890
\pm 260$)~km~s$^{-1}$ centred on systemic redshifts of $z=0.4216 \pm
0.0007$ and $z=0.4328 \pm 0.0008$, respectively. To investigate the
spatial distribution of the galaxies from the two components, we split
the sample at $z{=}0.4268$. Fig.~\ref{fig:overlay2} illustrates the
correlation between the spectroscopic redshifts of galaxies and their
positions on the sky, in which yellow and green circles are the
cluster galaxies at $z>0.4268$ and $z<0.4268$, respectively, red
circles are the multiple systems, and blue circles are the members of
A\,3192. Galaxies in the NE of \source\ tend to have higher redshifts
than those in the SW region.

\subsection{Strong gravitational lensing}\label{sec:lens_model}
\subsubsection{Strong-lensing features}

\begin{figure}
  \begin{center}
    \includegraphics[width=2.5cm]{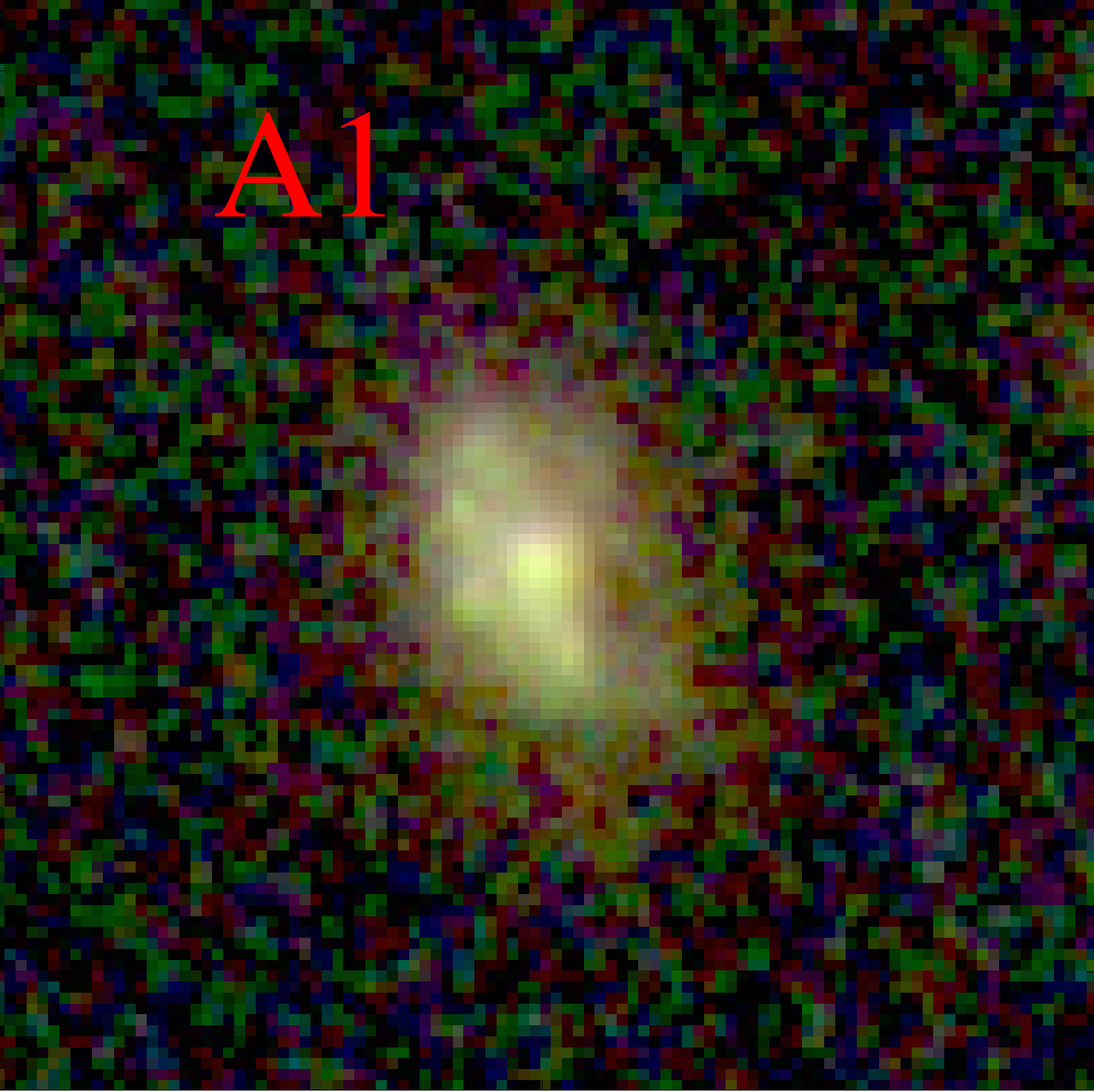}\mbox{}
    \includegraphics[width=2.5cm]{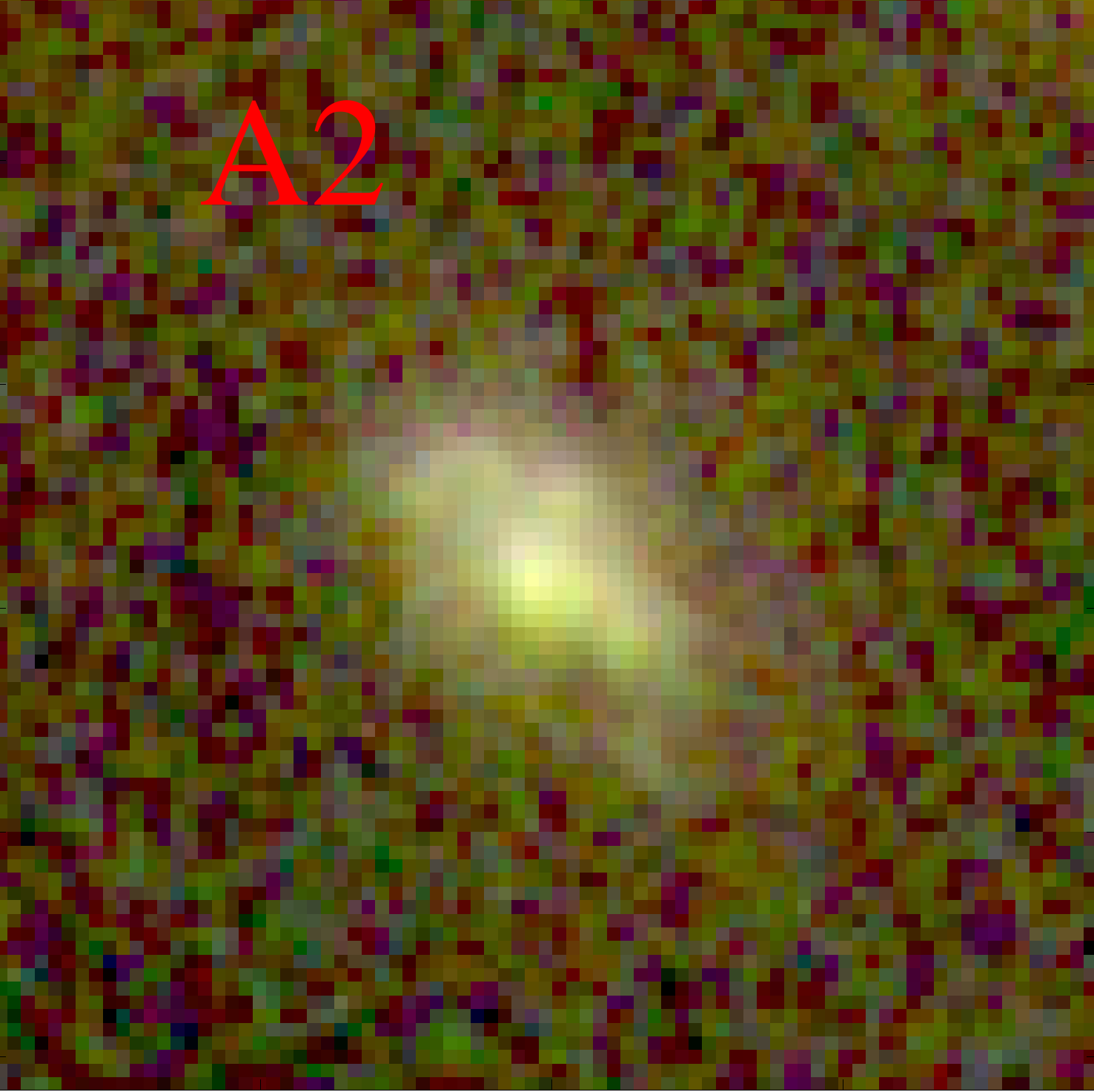}\mbox{}
    \includegraphics[width=2.5cm]{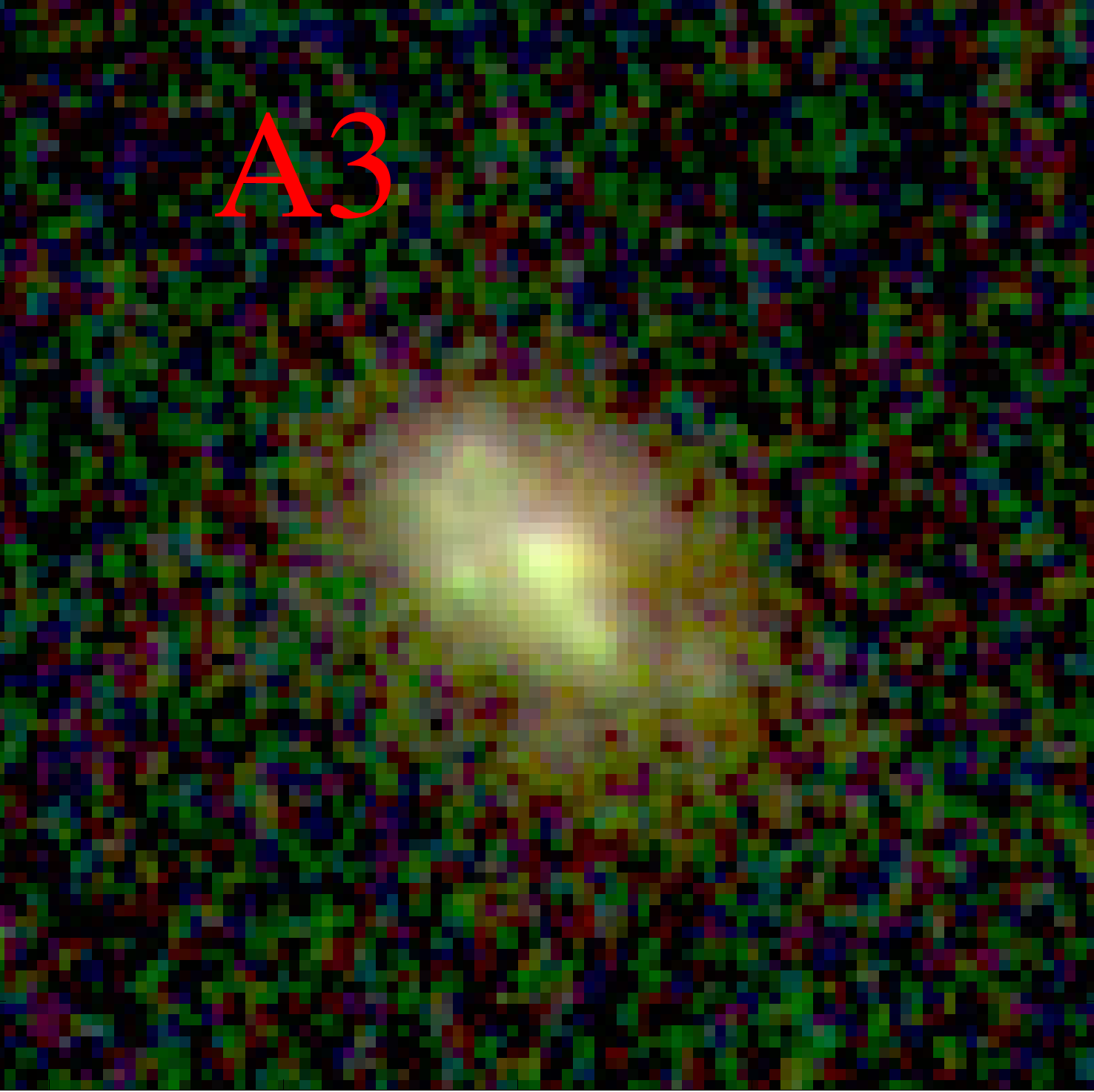}\mbox{}\\[2mm]
    \includegraphics[width=5.1cm]{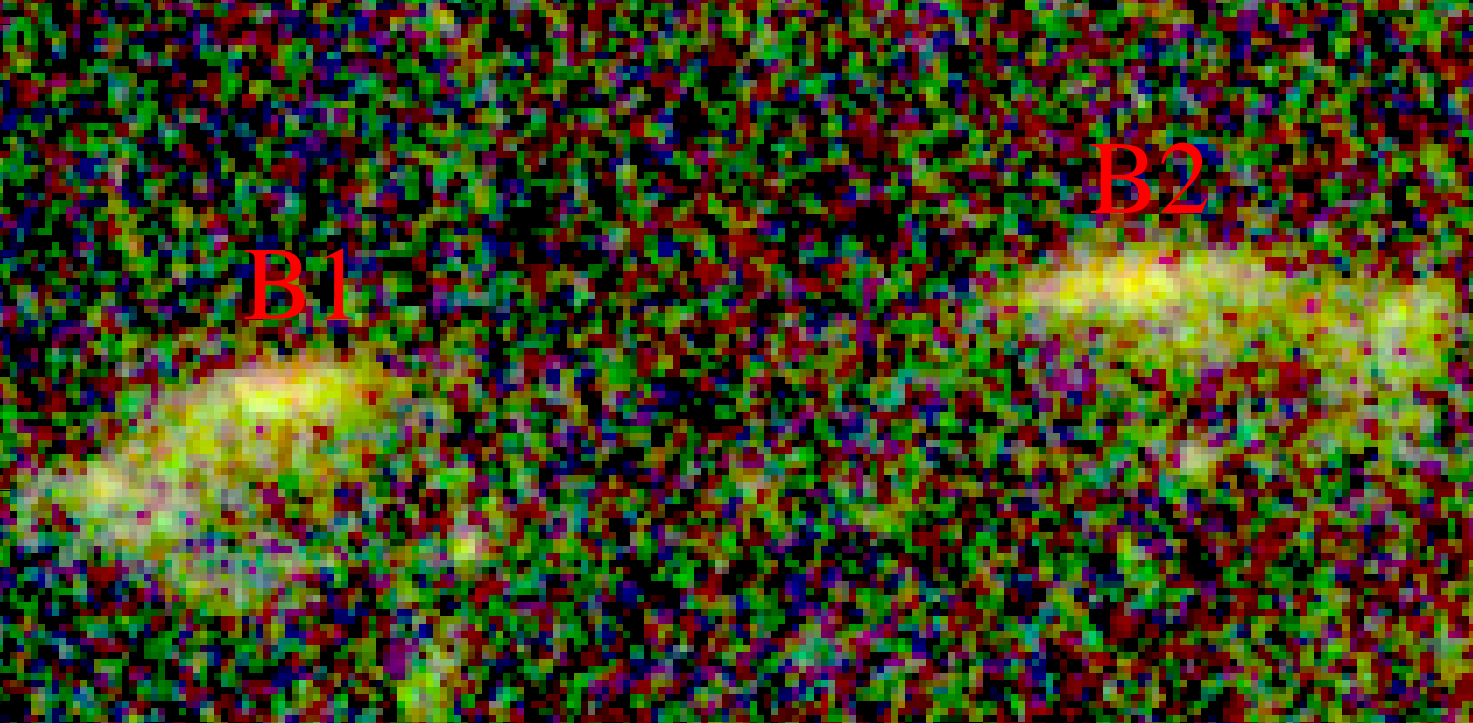}\mbox{}
    \includegraphics[width=2.5cm]{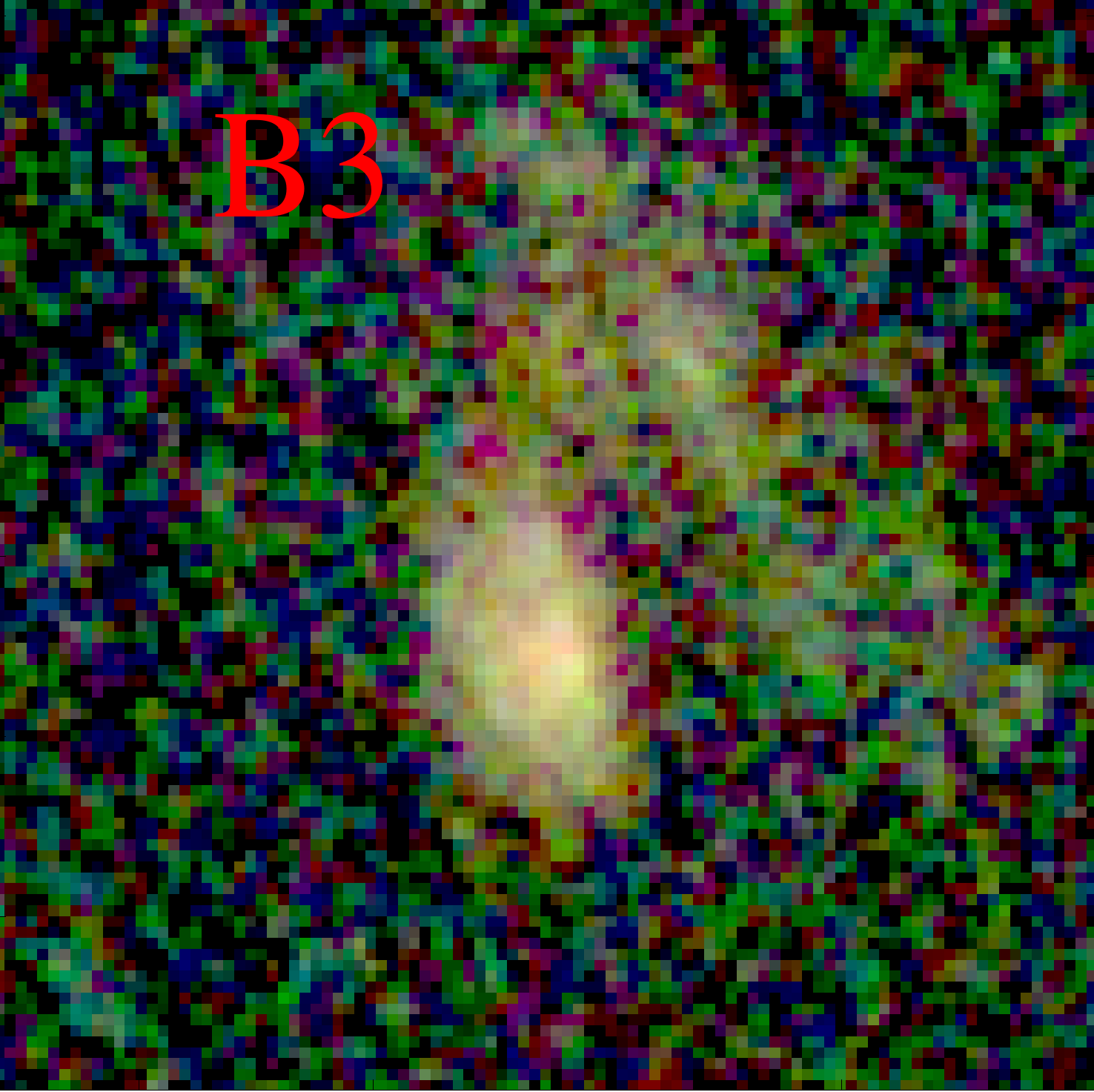}\mbox{}\\[2mm]
    \includegraphics[width=2.5cm]{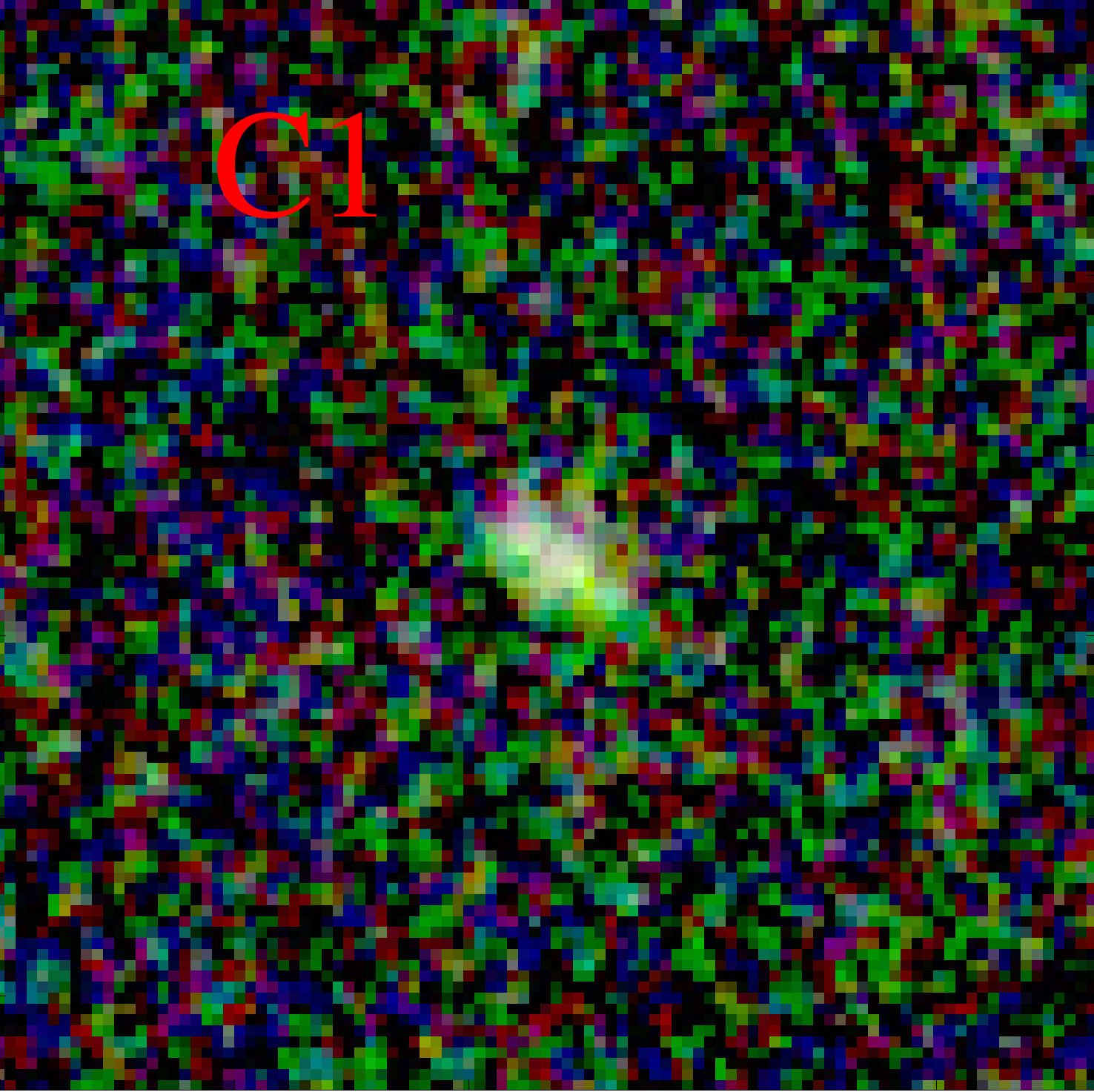}\mbox{}
    \includegraphics[width=2.5cm]{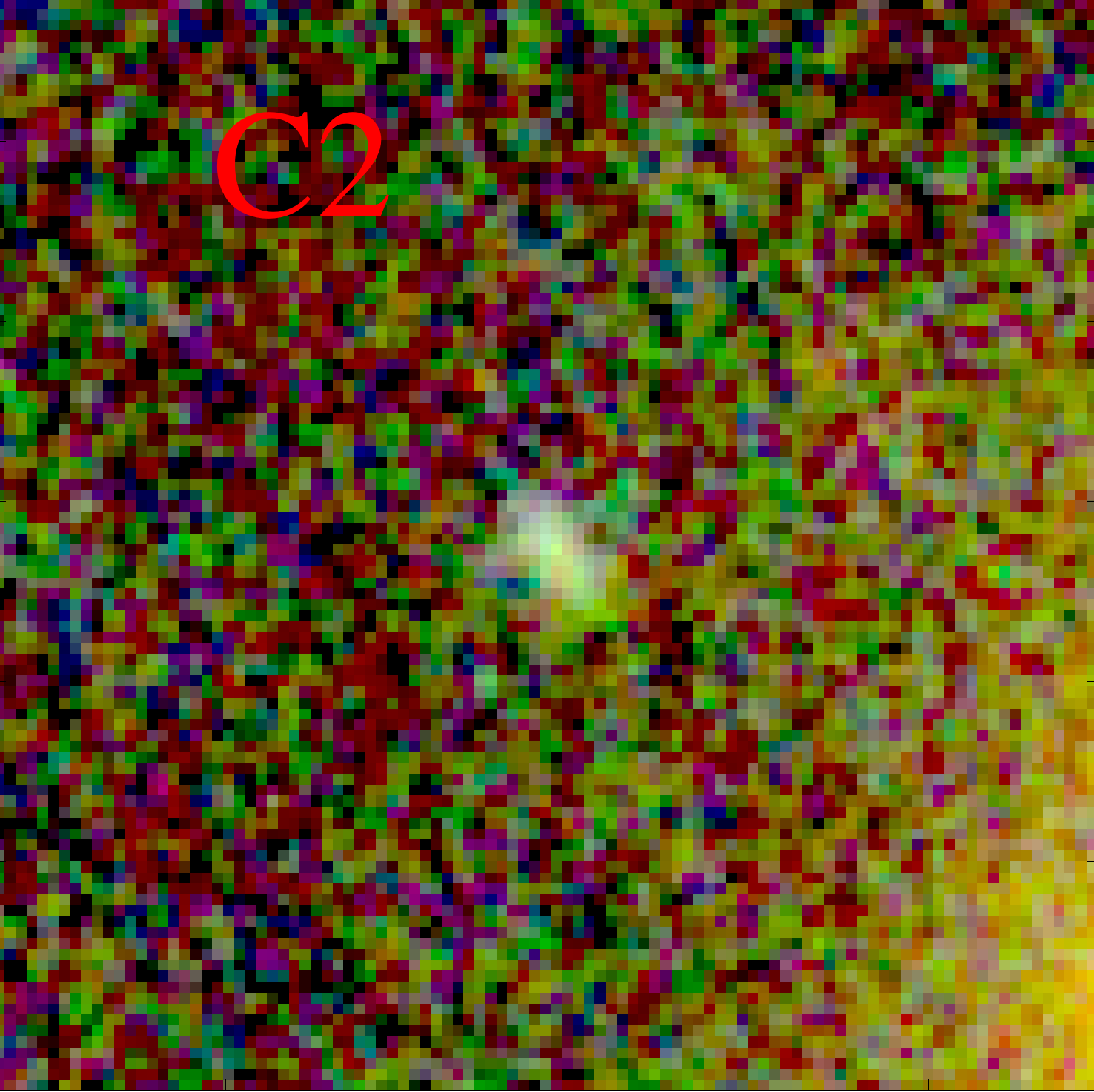}\mbox{}
    \includegraphics[width=2.5cm]{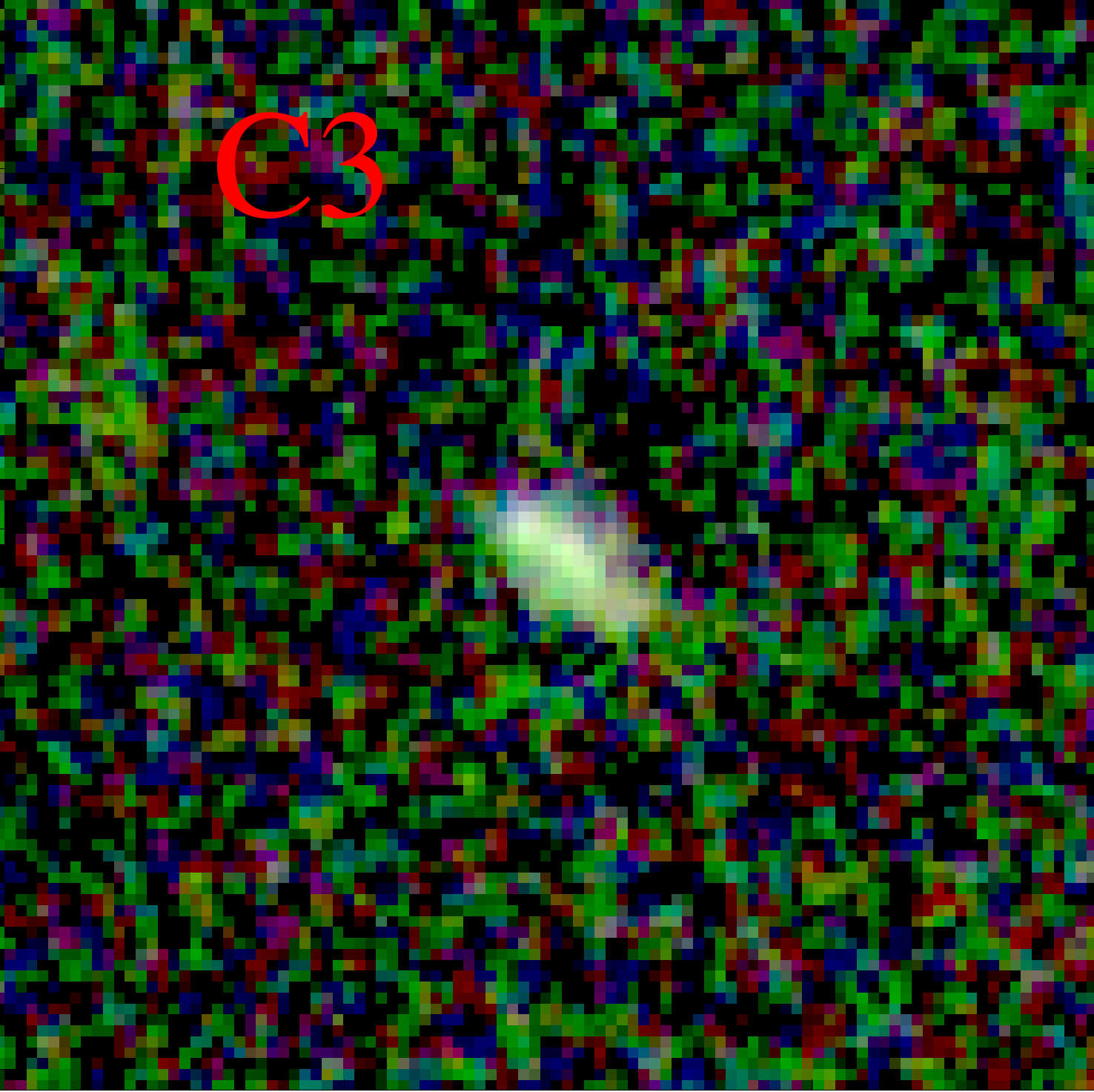}\mbox{}\\[2mm]
    \includegraphics[width=2.5cm]{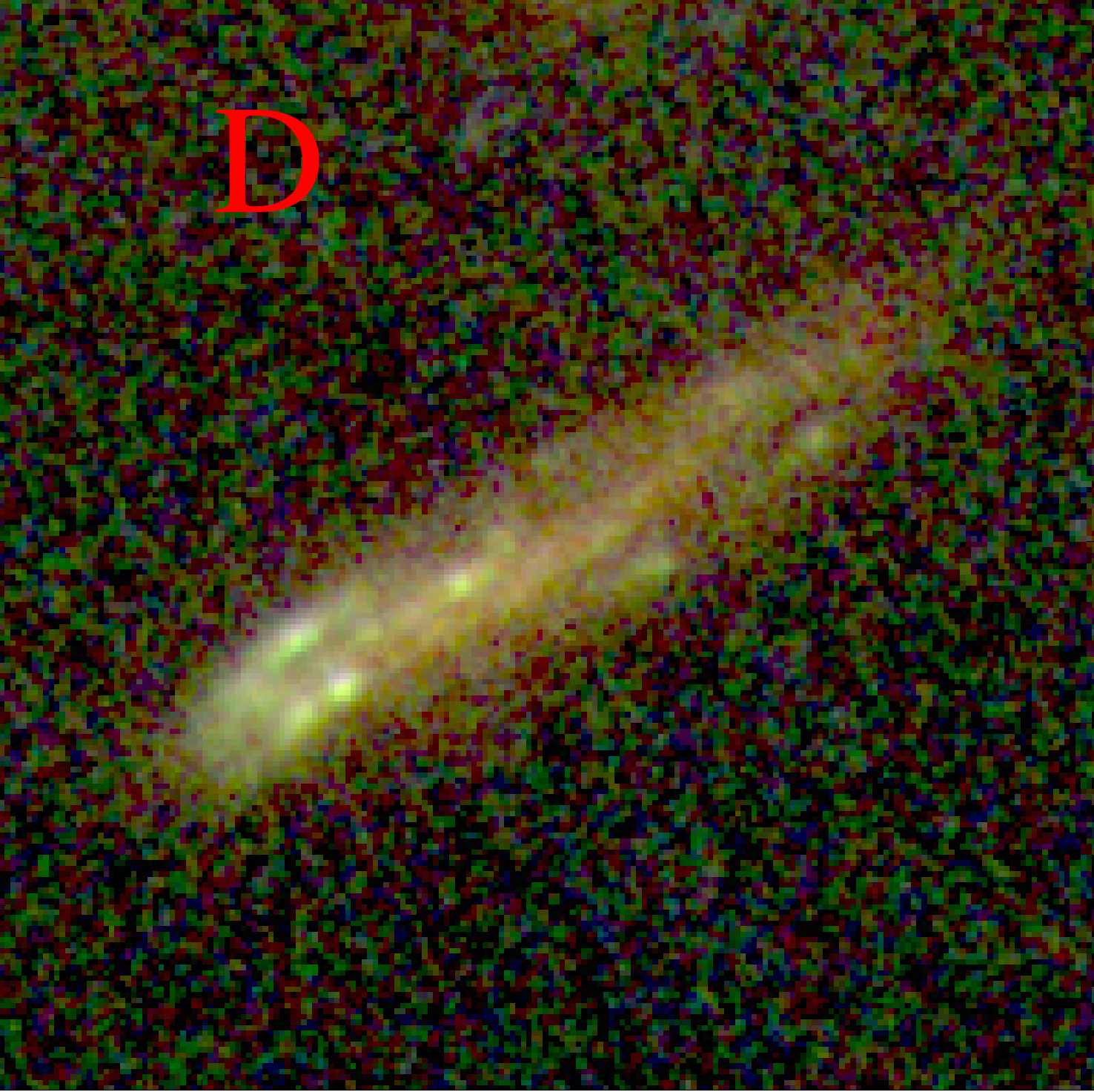}
    \caption{Close-ups of all strong-lensing features, i.e., the
      triples A, B, and C, and the single arc D
      (labels as in Fig.~\ref{fig:overlay2}).}
    \label{fig:multiples}
  \end{center}
\end{figure}

From the HST/ACS image shown in Fig.~\ref{fig:overlay2} we identify
three multiple-image systems (A, B, C) from their matching colours and
morphologies, as well as one extended arc-shaped source (D). Close-ups
of all images are shown in Fig.~\ref{fig:multiples}. Spectroscopic
redshifts were obtained for A1 and B1 with Keck-I/LRIS as detailed in
Section~\ref{sec:redshifts}. System B was independently proposed as a
multiple-image set (but not spectroscopically confirmed) by
\citet[][their system A1]{2012ApJ...748L..23H}.

System A ($z{=}1.262$) has an irregular morphology, the features of
which are readily recognisable in all three images
(Fig.~\ref{fig:multiples}). System B is a clear symmetric pair with a
spectroscopic redshift of $z{=}3.07$, for which our best lensing model
(described in detail below) predicts a third, fainter, counter-image
B3 at $\alpha{=}03\;58\;50.1$, $\delta{=}{-29}\;55\;23$ (J2000). This
position is quite different from the one predicted by
\citet{2012ApJ...748L..23H}, which is rejected at more than $10\sigma$
confidence by our improved model.  System C is a faint, blue, and
compact system of three images for which we could not secure a
spectroscopic redshift. However, it is fully compatible with our best
lensing model. Finally, we measured a spectroscopic redshift of
$z{=}1.537$ for the bright arc D; from our model, we expect no
counter-images and none are observed.

\begin{figure}
  \begin{center}
    \includegraphics[width=8.5cm]{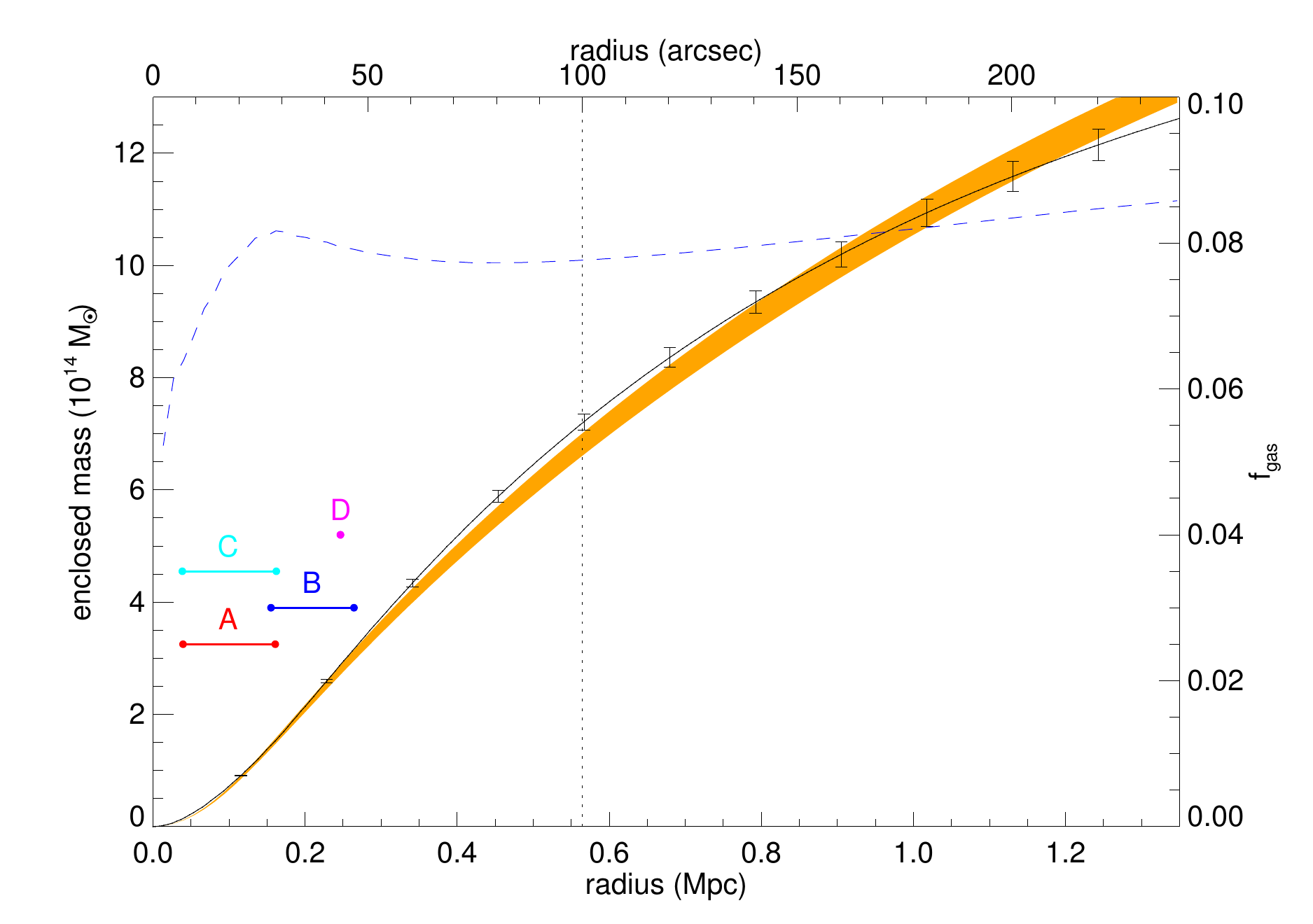}
    \caption{2D mass profile derived from the strong-lensing
      constraints. The vertical line marks the approximate radius
      beyond which the lens model should be considered an
      extrapolation; horizontal lines show the radial ranges of
      individual strong-lensing features, labelled as in
      Fig.~\ref{fig:overlay2}. The 2D gas mass profile from
      Fig.~\ref{fig:mass} is shown in orange, scaled by an assumed gas
      mass fraction of 8.2\% to match the total mass from our lensing
      analysis. The dashed blue line, finally, shows the gas mass
      fraction required for perfect agreement between the lensing and
      X-ray estimates of the total mass at all radii. }
    \label{fig:lens_mass}
  \end{center}
\end{figure}

\subsubsection{Mass model}

We use the strong-lensing features described in the previous section
to constrain a parametric mass model for the cluster using the
Lenstool\footnote{Publicly available at
  http://projets.oamp.fr/projects/lenstool} software
\citep{2007NJPh....9..447J}. Both very extended (cluster-scale) and
compact (galaxy-scale) mass components are included and parameterised
as pseudo-isothermal elliptical mass distributions (PIEMD), similarly
to the approach taken by, e.g., \citet{2007ApJ...668..643L} and
\citet{2009A&A...498...37R}. The parameters of all galaxy-scale mass
distributions are fixed based on their properties as compiled in our
photometric catalog of cluster members, i.e., we use the geometric
parameters of their light profiles and assume that light traces mass.

We start the modeling process with a single cluster-scale component,
allowed to move freely across the ACS field of view. This model fails
to reproduce the geometry of system A whose three images are almost
perfectly aligned.  Since such a configuration is typical of a
``saddle'' between two massive large-scale components, we next attempt
to satisfy the lensing constraints with a model comprised of two
cluster-scale halos whose positions are allowed to vary within
15\arcsec of each of the two main light peaks (1 and 2,
Fig.~\ref{fig:overlay2}). This model reproduces all three
multiple-image systems A, B, and C with a root-mean-square (rms)
positional uncertainty of less than $0.5\arcsec$. Image D is predicted
to be a single arc. The model also predicts a counter-image of the
image pair B1/B2 within a small region which is indeed found to
contain a plausible candidate (B3 in
Fig.~\ref{fig:multiples}). Whether or not this third image is included
as an additional constraint does, however, not change the best-fit
model. The best-fit parameters of our strong-lensing mass model of
\source\ are summarised in Table \ref{tab:lensmodel}. Although the
strong-lensing features in \source\ primarily constrain the mass
distribution on the whole, our model strongly prefers an inequitable
mass distribution between the two cluster-sized halos, with the NE
component being the more massive one.

For completeness' sake we note that \citet{2012ApJ...748L..23H}
propose an additional pair of images, located between our
multiple-image sets B and C, as another strong-lensing system. Lacking
spectroscopic redshifts, these two images do not add any additional
constraints to the model. Moreover, their morphologies and colours do
not fully agree, and we are unable to identify a plausible candidate
for the required counter-image. For these reasons we consider this
identification questionable and choose not to include this system in
our model.

Fig.~\ref{fig:lens_mass} shows the lensing-mass profile for the entire
system, centred at the same location as the X-ray gas mass profile
presented in Fig.~\ref{fig:mass}. Dividing the gas mass profile by the
profile of the total lensing mass yields the gas mass fraction as a
function of radius, shown as the dashed blue line in
Fig.~\ref{fig:lens_mass}. At small radii the resulting gas mass
fractions have little physical meaning since both profiles are centred
on a location that falls between the two cluster components. At larger
radii, the gas mass fraction is approximately constant, rising only
slightly from 8 to 8.5\%. This value implies a baryon fraction of
about 9\% which is at the low end of the observed range and possibly
an effect of the ongoing merger activity in
\source\ \citep{2012ApJ...748..120D}.

\begin{figure*}
  \begin{center}
    \includegraphics[width=17cm]{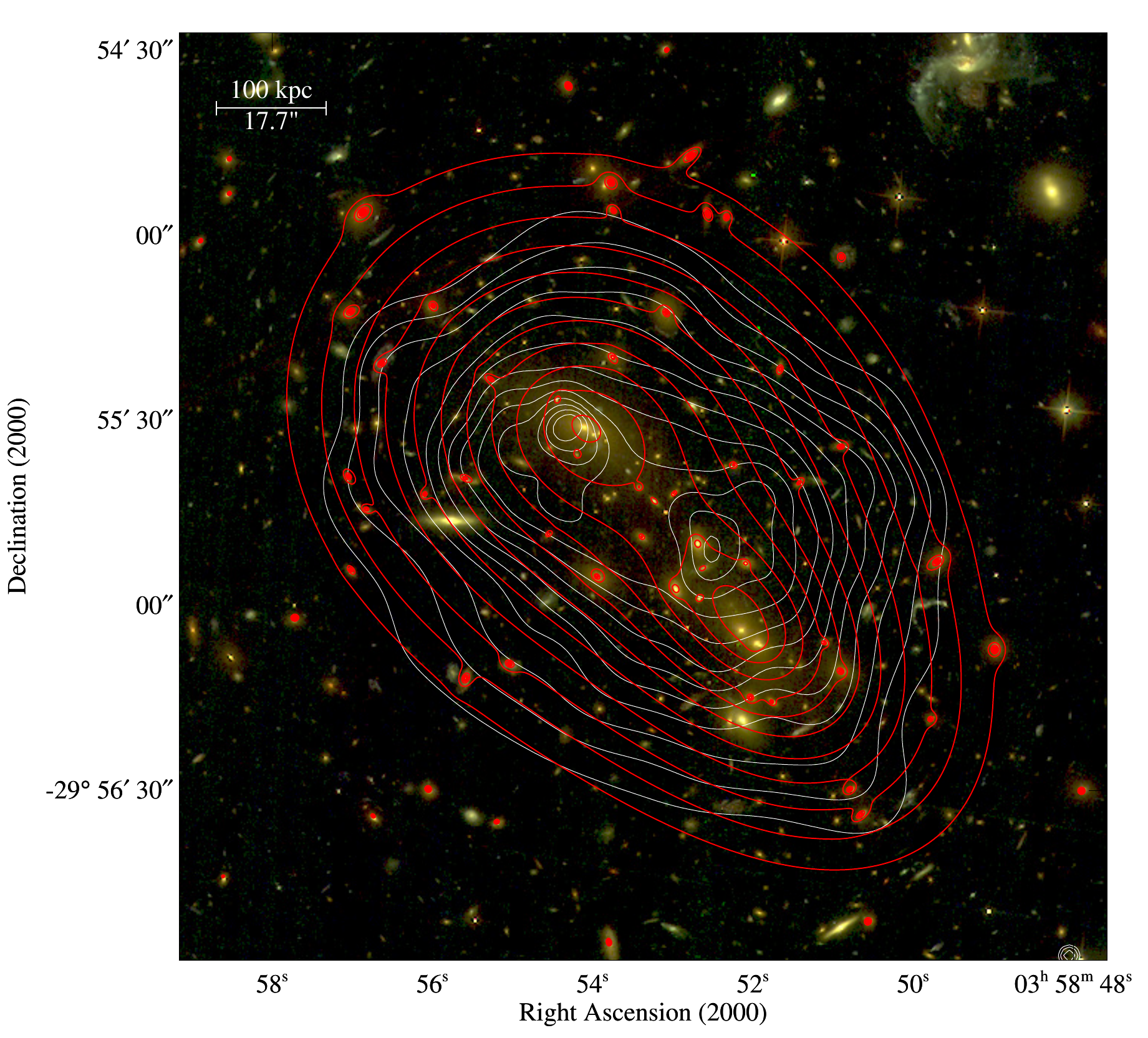}
    \caption{Isointensity contours of the adaptively smoothed X-ray
      emission from \source\ as observed with Chandra/ACIS-I (white)
      and contours of the total gravitational mass (red) as derived by
      strong lensing analysis overlaid on the HST colour image (blue:
      F435W, green: F606W, red: F814W).}
    \label{fig:overlay3}
  \end{center}
\end{figure*}

Fig.~\ref{fig:overlay3} shows contours of the total mass as determined
from our strong-lensing analysis overlaid on the data shown before in
Fig.~\ref{fig:overlay}. The offset between the X-ray and mass contours
seen in Fig.~\ref{fig:overlay3} for the SW sub-cluster will be discussed
in Section 5.2.

\begin{table*}
\begin{tabular}{lccccccc}
\hline
Potential & X            & Y            & $e$           & $\theta$ & r$_{\rm core}$ & $\sigma$ & r$_{\rm cut}$\\
          &  [\arcsec]   & [\arcsec]    &               & [deg]     &     [kpc]     &  km s$^{-1}$      & [kpc] \\[2mm]
NE & $-$2.1$\pm$1.1 & 0.65$\pm$1.0 & 0.14$\pm$0.06 & 157$\pm$5 & $99.3\pm6.2$ & 1110$\pm$32 & [1000] \\
SW & 26.0$\pm$0.5 & $-$30.9$\pm$0.6& 0.47$\pm$0.04 & 126$\pm$2 & $48.3\pm5.2$ & 813$\pm$30 & [1000] \\
\hline
\end{tabular}
\caption{\label{tab:lensmodel}Parameters of the two PIEMD potentials
  used to model the mass distribution. From left to right: coordinates
  of centre (measured in $\arcsec$ from the BCG of the NE component),
  ellipticity, position angle, core radius, velocity dispersion,
  truncation radius (fixed at 1 Mpc). A redshift of 1.65$\pm$0.02 is
  assumed for System C.}
\end{table*}

\section{Discussion}
\label{sec:discussion}

\subsection{Cluster scaling relations}

\begin{figure}
  \begin{center}
    \includegraphics[width=7cm]{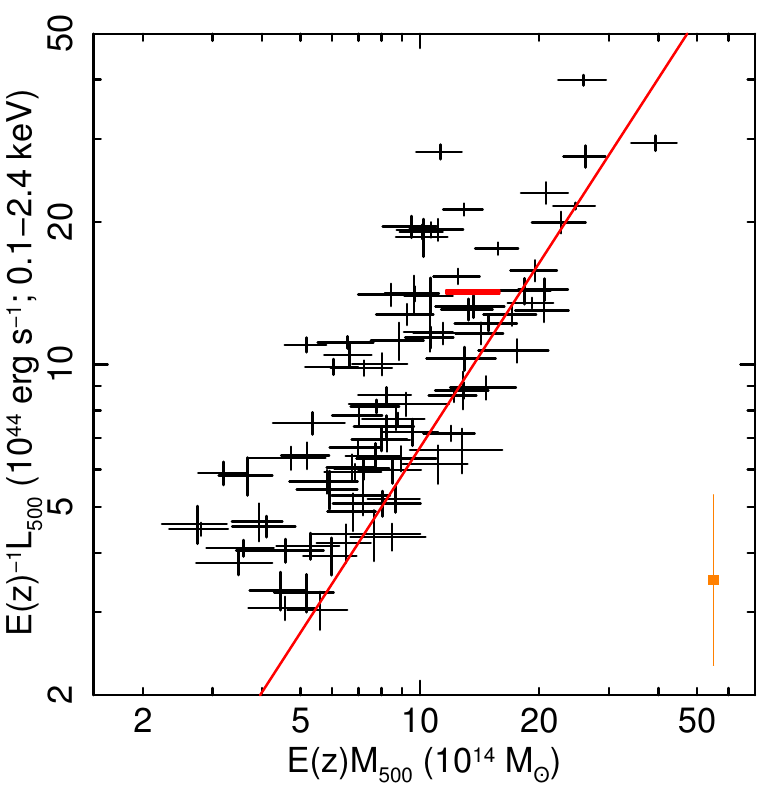}
    \includegraphics[width=7cm]{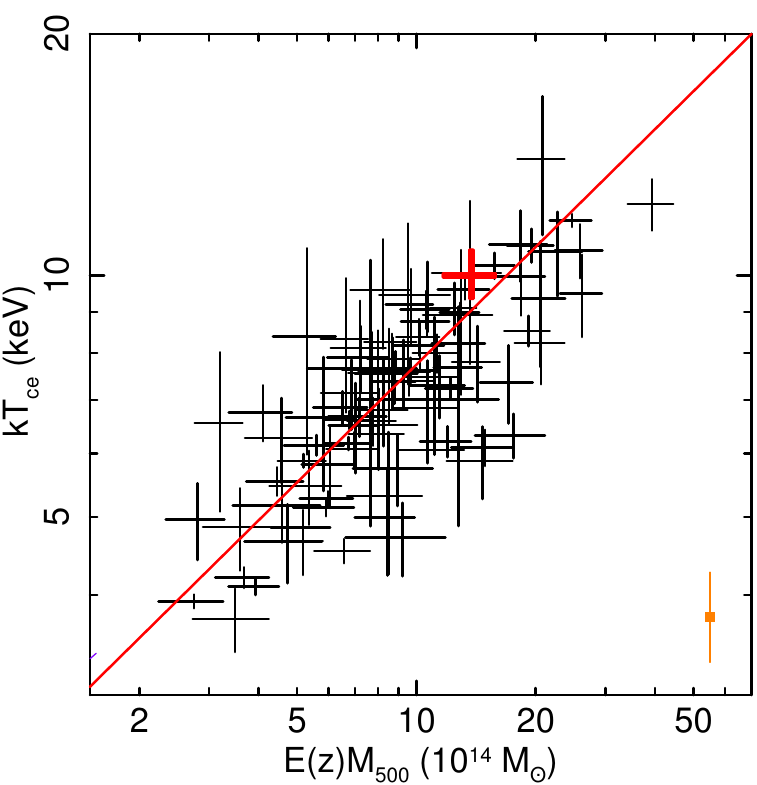}
    \includegraphics[width=7cm]{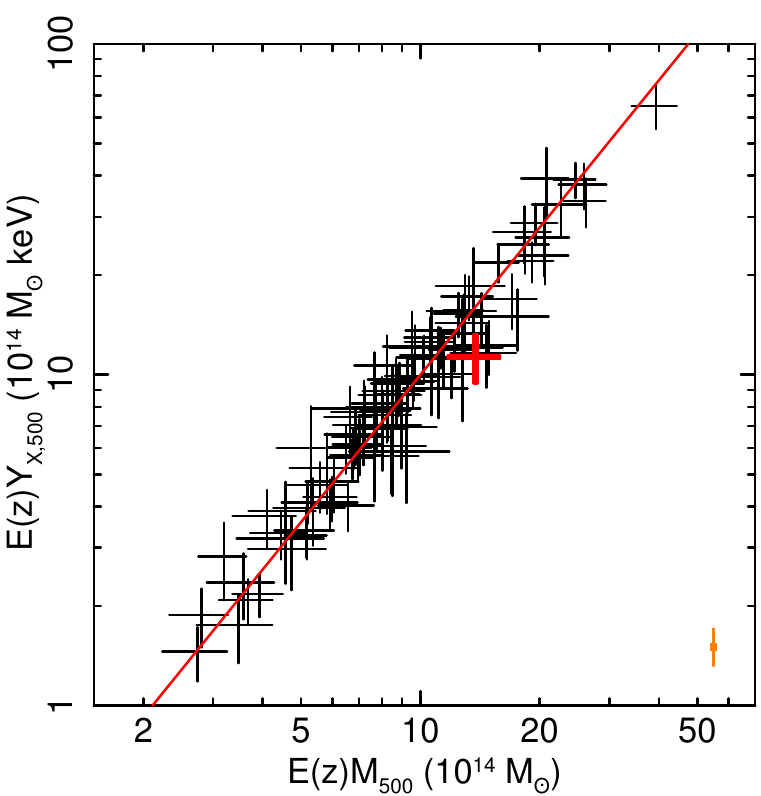}
    \caption{Cluster scaling relations from
      \citet{2010MNRAS.406.1773M} with data point for
      \source\ overlaid in red. The fiducial data point in the lower
      right of each plot illustrates the $1\sigma$ scatter of the
      respective relationship. From top to bottom: $L_X-M$, k$T-M$,
      and $Y_X-M$. All quantities are computed within $r_{\rm 500}$;
      k$T$ is the core-excised gas temperature.}
    \label{fig:scaling}
  \end{center}
\end{figure}

To further assess the effects of merger activity within \source\ we
compare the system's global X-ray properties to the averages provided
by the $L_{\rm X}-M$, k$T-M$, and $Y_X-M$ scaling relations of
\citet{2010MNRAS.406.1773M}. Our estimate of the total mass of $M^{\rm
  3D}(<r_{500}) = (1.12 \pm 0.18) \times 10^{15} M_{\odot}$, the
global 0.1--2.4 keV luminosity $L_X(<r_{500})$ of $1.76 \times
10^{45}$ erg s$^{-1}$, and the core-excised temperature of 10.0 keV
fall within the $2\sigma$ scatter of all three scaling
relations. Fig.~\ref{fig:scaling} shows both the X-ray luminosity and
the global gas temperature of \source\ to be slightly high compared to
the best-fit predictions based on the cluster's mass. This is
consistent with a mild boost in both quantities as a result of the
ongoing merger. A more pronounced increase in temperature and
luminosity as seen in numerical simulations of head-on cluster
collisions \citep[e.g.][]{2002ApJ...577..579R} is not expected for a
merger proceeding at non-zero impact parameter like \source.

\subsection{Three-dimensional merger dynamics and geometry}

Cold fronts originate when regions of cool, dense gas move
subsonically through a less-dense and hotter medium. We refer to
\citet{2007PhR...443....1M} for an excellent review of the related
physical and observational phenomena. In brief, depending on the
impact parameter of the encounter and the viewing geometry, we may
observe a single, or two cold fronts. Dual cold fronts, one much more
compact and pronounced than the other, are observed when ``sloshing''
is induced in the cool core of a cluster as the result of the passage
of a massive perturber, which is in general another cluster (see
\citealt{CJ} for a study of one such case).

The presence of two likely cold fronts in \source, discussed in
Section~\ref{sec:coldfront}, strongly suggests that we are observing a
massive cluster merger after its first core passage.  This insight has
important implications for our interpretation of the three-dimensional
merger trajectory and history, as detailed in the following.

Our imaging and spectroscopic analysis at both optical and X-ray
wavelengths confirms the primary hypothesis of ME12 that \source\ is
an active merger of (at least) two sub-clusters. Spectroscopic galaxy
redshifts within the area shown in Fig.~\ref{fig:overlay} show a
bimodal distribution centred at $z{=}0.422$ and $z{=}0.433$
(Fig.~\ref{fig:z_hist}), and a small population of foreground galaxies
at $z{=}0.168$ (Fig.~\ref{fig:overlay2} and
Tables~\ref{lris_redshift},\ref{lris_redshift2}).  In conjunction with
our Chandra data, which establish conclusively that next to all of the
detected X-ray emission originates from \source, this firmly rules out
the possibility that the superposition of A\,3192 ($z{=}0.168$)
constitutes a significant source of contamination or confusion.

As for the bimodal distribution of galaxy redshifts in \source, we
note a pronounced correlation between radial velocity and position on
the sky (Fig.~\ref{fig:overlay2}), such that galaxies in the NE
sub-cluster feature systematically higher redshifts than those in the
SW system.  This difference in redshift could be attributed either to
Hubble expansion (i.e., the NE sub-cluster is more distant) or to
peculiar velocities due to infall. We can rule out the first scenario
since it implies a distance between the two sub-clusters of more than
10 Mpc along the line of sight, which is in conflict with clear signs
of physical interaction (cold fronts) detected in our X-ray analysis
of the system (Section~\ref{sec:coldfront}). Adopting the alternative
explanation that the bimodal distribution shown in
Fig.~\ref{fig:z_hist} is due to peculiar velocities, we conclude that
the two sub-clusters are near their first passage, and are moving past
each other with a relative speed of roughly 1500 km s$^{-1}$ along the
line of sight. Applying Occam's razor, this scenario is also favoured
as an explanation for the small projected separation of the cluster
cores ($<$200 kpc at $z{=}0.428$). We conclude that the merger axis is
significantly, but not fully, aligned with our line of sight.

\begin{figure}
    \includegraphics[width=8cm]{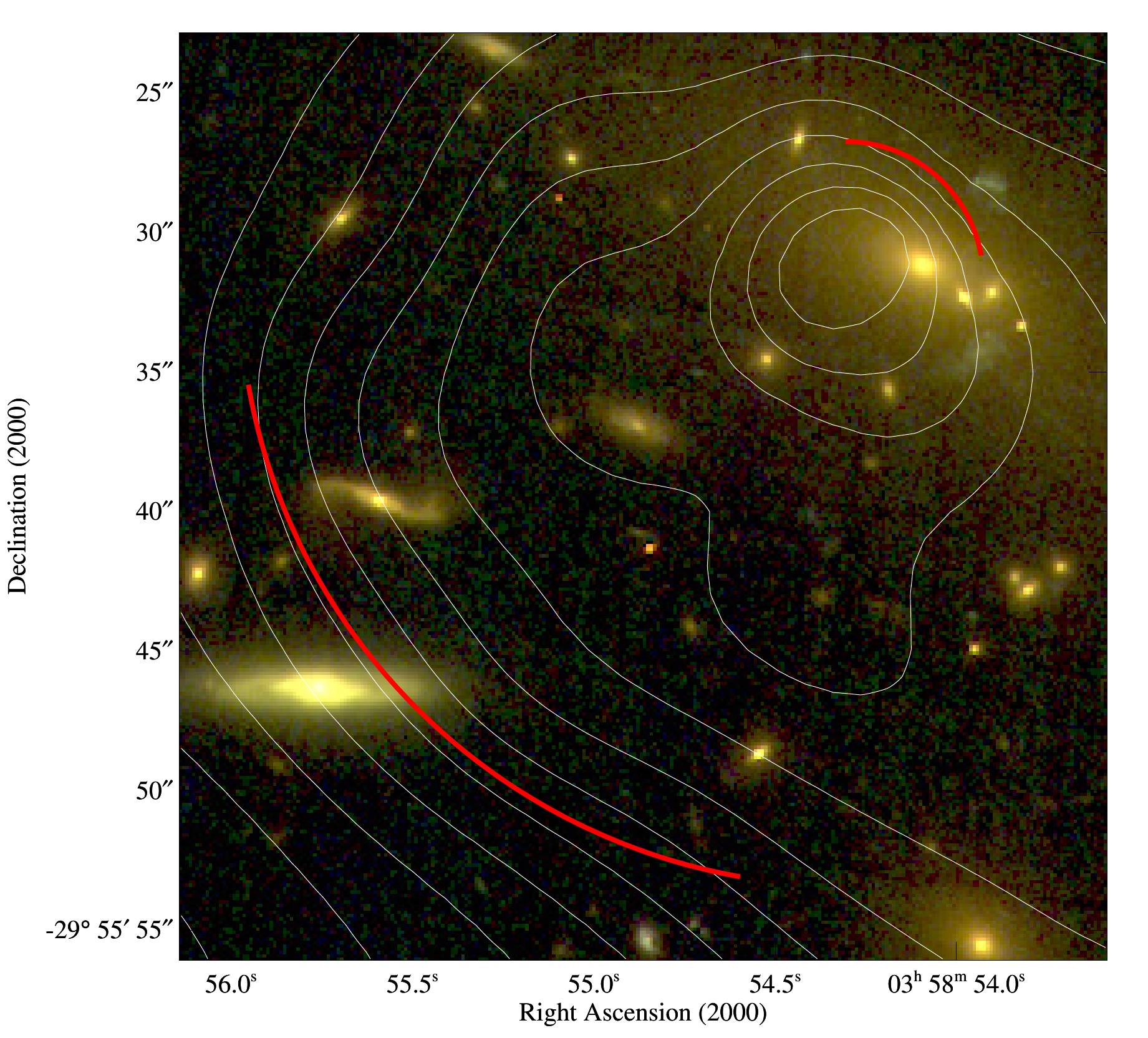}
    \caption{Close-up view of the core of the NE component of
      \source\ as shown in Fig.~\ref{fig:overlay}. Note the
      well-aligned two cold fronts (marked by the red arcs) and the
      offset of the NE sub-cluster's BCG from the X-ray peak.}
    \label{fig:overlay_zoom}
\end{figure}

In order to further constrain the three-dimensional merger geometry we
consider the location and alignment of the two cold fronts and the
relative locations of gas and galaxies in the vicinity of the core of
the NE sub-cluster. As shown in Fig.~\ref{fig:overlay_zoom}, both the
alignment of the two cold fronts and the offset of the NE
sub-cluster's BCG from the associated X-ray peak imply a projected
direction of motion along a SE-NW axis. Such a motion is fully
consistent with gas sloshing induced by the passage of a second
cluster to the north-west of the cool core (the side of the more
pronounced cold front), proceeding almost along our line of sight. The
passage cannot have occurred exactly along our line of sight because
the cold fronts are not aligned with the line connecting the two
sub-clusters of \source. This proposed trajectory of the SW
sub-cluster is supported also by the results of our strong-lensing
analysis: the offset between the centre of the gravitational mass of
the SW cluster and the associated X-ray peak
(Fig.~\ref{fig:overlay3}), most likely induced by the interaction of
the two cluster components during their close passage, points to an
origin of the SW component that lies just north of the dominant, NE
cluster.

\begin{figure}
  \begin{center}
    \includegraphics[width=8cm]{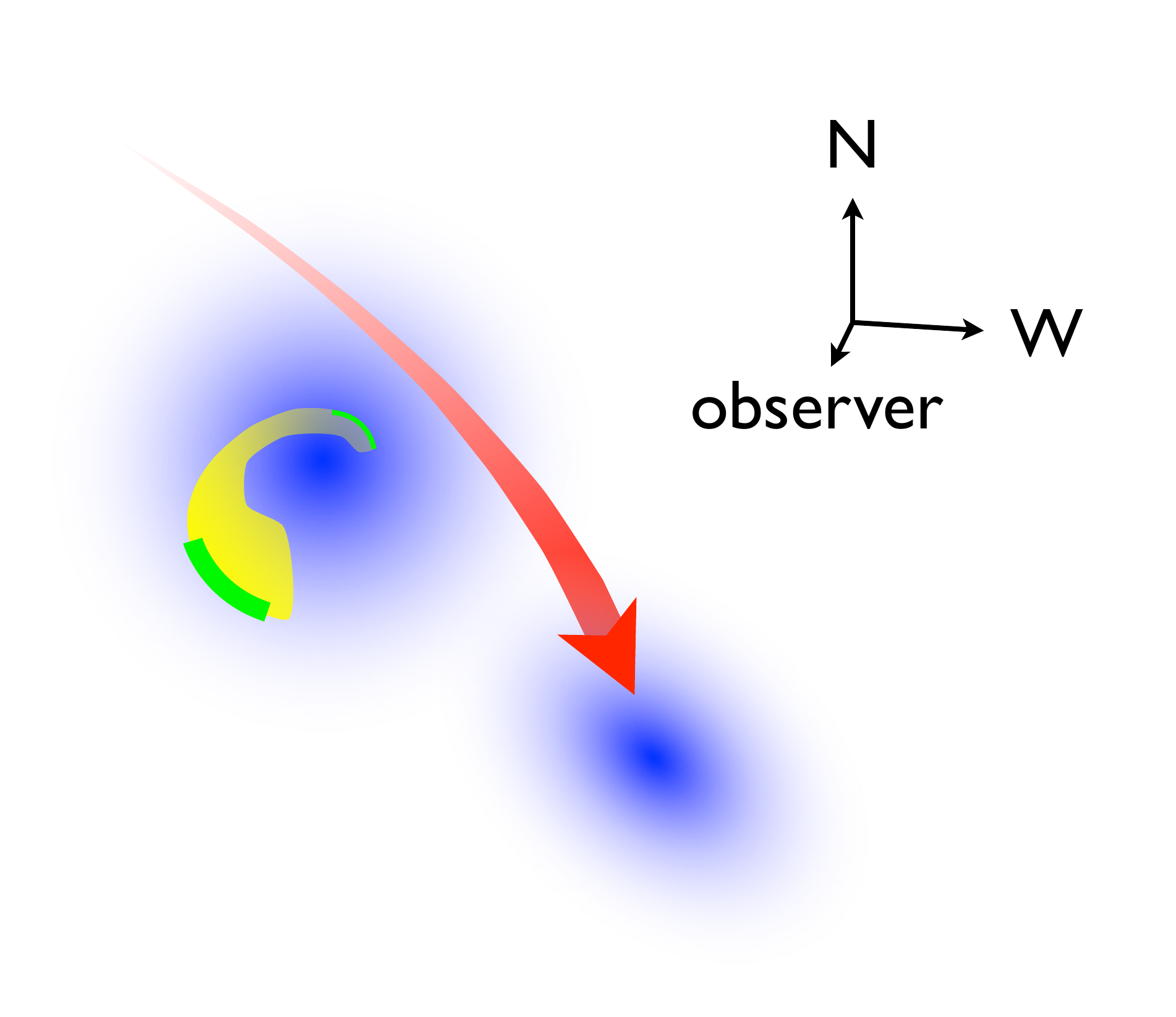}
    \caption{Schematic view of the merger geometry and trajectory. The
      SW component of \source\ passed north, and mainly behind, the NE
      component at a small but significant impact parameter (red
      trajectory), inducing sloshing of the cold core of the NE
      cluster. Along our line of sight, the spiral distribution of
      cool gas (yellow) created in this process is most clearly
      discernible where its edges are viewed in projection -- these
      are the visible cold fronts (green).}
    \label{fig:3dsketch}
  \end{center}
\end{figure}

We thus conclude that \source\ is observed after the first core
passage of a merger at non-zero impact parameter along an axis that is
almost aligned with our line of sight, and that the trajectory of the
incoming sub-cluster has changed from a south-westerly to a more
southern direction (in projection), such that the secondary
sub-cluster is now observed to the SW of the dominant NE cluster.  As
a result, the line connecting the two sub-clusters as observed is
neither the merger axis nor does it indicate the current direction of
motion. A sketch illustrating the proposed merger geometry and
dynamics is shown in Fig~\ref{fig:3dsketch}.

Although \source\ is primarily a merger of just the two sub-clusters
discussed so far, we find evidence at both X-ray and optical
wavelengths that adds further complexity to this scenario. As is
apparent from Fig.~\ref{fig:overlay}, the SW sub-cluster lacks an
obvious BCG. Instead, several bright ellipticals are observed near the
cluster core, and the location of none of them coincides with the peak
of the X-ray emission. In addition, our spectroscopic analysis of the
X-ray emission in this region (Fig.~\ref{fig:map}) finds compelling
evidence of a mix of cool and very hot gas ($\sim$16 keV). Although
the gas dynamics and temperature cannot be more accurately constrained
with the photon statistics of the data in hand, we propose that the SW
sub-cluster is itself highly disturbed, most likely as the result of
an ongoing merger at high inclination angle.

We conclude that \source\ is in fact not a BHOM but most likely a
double merger, one of which proceeds almost head-on but along our line
of sight, while the other features a significant impact parameter and again
a significant inclination angle with respect to the plane of the sky. 

\section{Summary}
\label{sec:summary}

We present an X-ray/optical analysis of \source\ based on observations
with Chandra, HST, and Keck/LRIS. We measure the system's global
properties (velocity dispersion, bolometric X-ray luminosity, gas
temperature, total gas mass, and total gravitational mass) as
$\sigma=1440^{+130}_{-110}\rm~km~s^{-1}$ from 41 galaxies (890 km
s$^{-1}$ when the relativistic equation is used),
$L_{X,\mathrm{bol}}(<r_{500}){=}4.24 \times 10^{45}$ erg s$^{-1}$, kT
= $9.55^{+0.58}_{-0.37}$ keV, $M^{\rm 3D}_{gas}(<r_{500}) = (9.18 \pm
1.45) \times 10^{13} M_{\odot}$, and $M^{\rm 3D}_{tot}(<r_{500}) =
(1.12 \pm 0.18) \times 10^{15} M_{\odot}$ (assuming a gas mass
fraction of 8.2\%).
  
Based on the systems's X-ray/optical morphology, we propose that
\source\ is an ongoing double merger.  The primary merger of the SW
and NE components features a significant impact parameter and a high
inclination angle with respect to the plane of the sky. From an
analysis of the offsets between collisional and collisionless matter
in \source\ as well as from the presence, location, and orientation of
two cold fronts near the core of the NE cluster, we conclude that the
SW sub-cluster passed behind and north of the NE sub-cluster, inducing
gas sloshing in the cool core of the latter. The lack of an obvious
BCG and the evidence of a mix of cool and very hot gas in the core
region of the SW component suggests that the passing cluster is itself
highly disturbed, most likely due to an ongoing merger at high
inclination angle. This scenario of a massive cluster passing and
merging with an even more massive one along the trajectory indicated
in Fig.~\ref{fig:3dsketch} is supported by the results of a
strong-lensing analysis of the system which also allows us to
determine the gas mass fraction of \source\ (8.2\%).

Finally, we show that the foreground galaxy group A\,3192 is
negligible compared to \source\ both in terms of total X-ray
luminosity and mass.

Our study thus confirms that \source\ is a highly disturbed ongoing
merger of two massive galaxy clusters. The three-dimensional geometry
of the encounter, however, does not support the hypothesis of ME12
that \source\ is a head-on, binary merger. Nonetheless, our study
establishes \source\ as a dynamically active, massive, post-collision
merger that could provide important insights into the impact of
cluster collisions on galaxy evolution in a high-density environment.

\section*{Acknowledgements}

LYH and HE gratefully acknowledge financial support from SAO grants
GO0-11140X and GO1-12172X, as well as from STScI grant GO-12313.  JR
is supported by the Marie Curie Career Integration Grant 294074. This
work made use of the data analysis packages CIAO, Chips, and, Sherpa
provided by the Chandra X-ray Center (CXC). We thank Chao-Ling Hung
and Matthew Zagursky for contributions to the HST and Keck/LRIS data
reduction, and the UH Time Allocation Committee for their support of
the groundbased follow-up observations for this study.

\appendix

\section{A\,3192} 

\begin{figure*}
  \begin{center}
    \includegraphics[width=16cm]{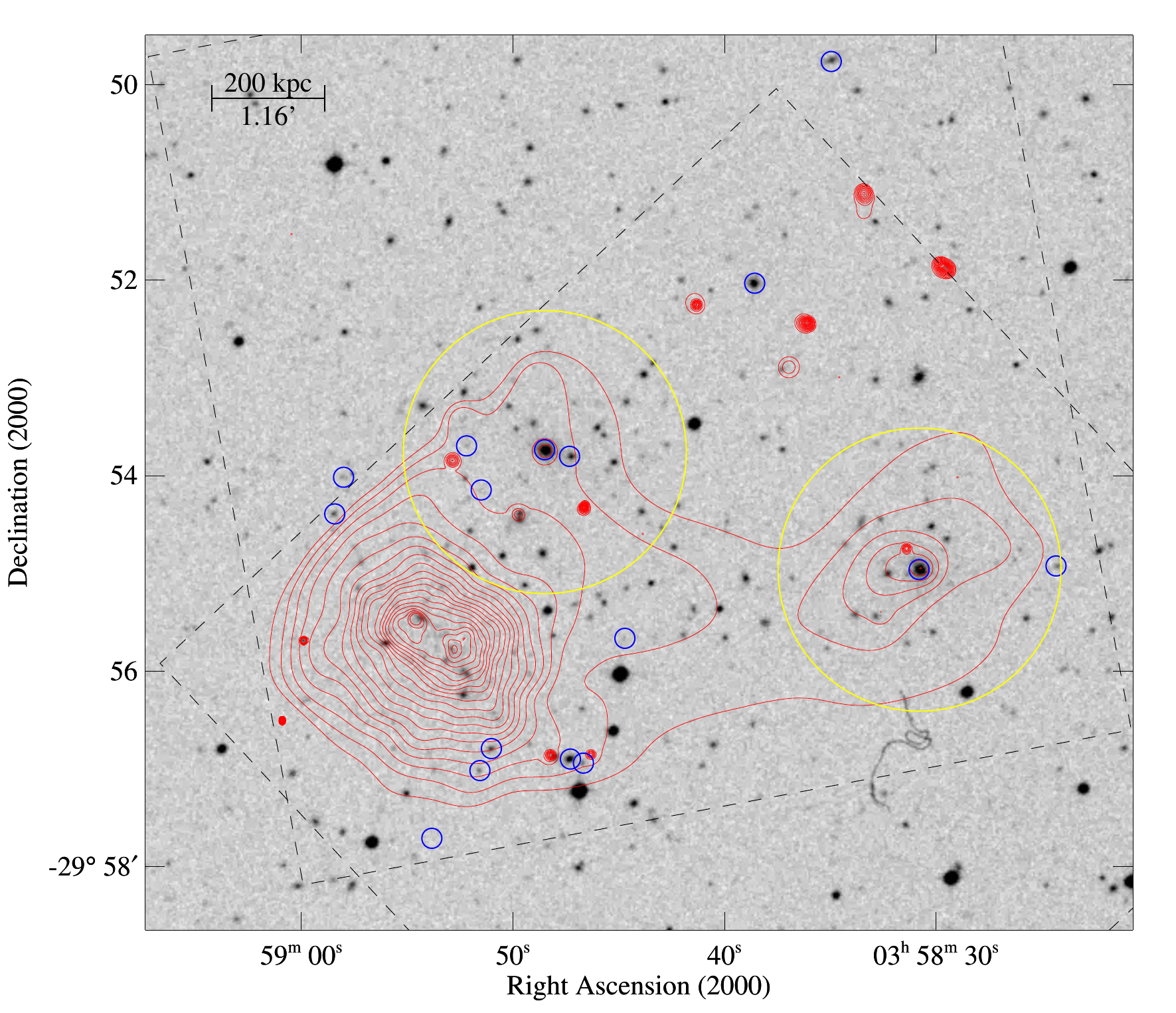}
    \caption{Isointensity contours of the adaptively smoothed X-ray
      emission observed with Chandra (ACIS-I) overlaid on the
      POSS2/UKST red image from the STScI Digitized Sky Survey. Blue
      circles mark the positions of the cluster galaxies of A\,3192
      determined by this work, \citet{2005AJ....130.2012W}, and
      \citet{2009A&A...499..357G}. Black dashed lines delineate the
      edges of the ACIS-I2 chip from the observation in October 2009
      (ObsID 11719; lower square) and November 2010 (ObsID 12300 and
      13194; upper square), respectively. Yellow circles show the
      apertures used to measure the X-ray flux and luminosity of
      A\,3192. The scale bar shown in the upper left assumes $z{=}
      0.168$. }
    \label{fig:overlaylarge}
  \end{center}
\end{figure*}

Our measurements of spectroscopic redshifts in the field of
\source\ identify ten galaxies with $z\sim 0.168$, i.e., members of
the foreground system A\,3192. X-ray emission from A\,3192 is detected
with Chandra (ACIS-I) to the north-west and west of
\source. Fig.~\ref{fig:overlaylarge} shows an overlay of the X-ray
emission on the POSS2/UKST red image from the STScI Digitized Sky
Survey, with blue circles marking members of A\,3192 identified by
this work, \citet{2005AJ....130.2012W}, and
\citet{2009A&A...499..357G}. The X-ray surface brightness distribution
of A\,3192 shows two peaks that coincide with two bright galaxies, and
a very weak feature connecting the two. We place two apertures centred
on the two bright galaxies to estimate the fluxes within r = 250 kpc
at z = 0.168 (87$''$) for the western and the north-eastern X-ray
peaks. The two apertures are shown as yellow circles in
Fig.~\ref{fig:overlaylarge}.

Since the photon statistics within the two apertures are insufficient
for spectral fitting, we convert the measured count rates to X-ray
fluxes and luminosities using in CIAO 4.4 and the spectral model
described in Section~\ref{sec:spectral}. We freeze the redshift at
0.168, the metal abundance at 0.3, and the absorption at the Galactic
value, and obtain an estimated plasma temperature iteratively from the
observed counts in the 0.1--2.4 keV band and the $L_X-{\rm k}T$
relation of \citet{2011A&A...535A.105E}. The unabsorbed bolometric
fluxes determined by this process for the two apertures are $(9.07 \pm
1.60)\times 10^{-14}$~erg~cm$^{-2}$~s$^{-1}$ and $(9.33 \pm 1.72)
\times 10^{-14}$~erg~cm$^{-2}$~s$^{-1}$ for the western and the
north-eastern apertures, respectively. The corresponding bolometric
luminosities in the A\,3192 rest frame are $(7.14 \pm 1.26)\times
10^{42}$~erg~s$^{-1}$ and $(7.34 \pm 1.36) \times
10^{42}$~erg~s$^{-1}$, and the estimated temperatures are 1.14 keV and
1.16 keV. The quoted uncertainties correspond to 1$\sigma$ confidence
and assume that the photon counts follow Poisson statistics. Since the
north-eastern aperture is contaminated by emission from the outskirt
of \source, the flux and luminosity of the NE component of A\,3192
should be considered an upper limit. For the quoted X-ray
luminosities, the $L_X-M$ relation of \citet{2011A&A...535A.105E}
predicts the masses of the two components of A\,3192 to be essentially
identical, at $2.75 \times 10^{13} M_{\odot}$ and $2.80 \times 10^{13}
M_{\odot}$.

\citet{2012ApJ...748L..23H} perform a weak-lensing analysis within the
field of their HST snapshot observation of \source\ and measure masses
in two apertures. The first aperture is centred on the BCG of
\source's NW sub-cluster, while the other is centred on the peak of
the luminosity-density map of likely members of A\,3192 selected by
colour from images obtained by supplementary groundbased
observations. This second position falls outside the ACS field of view
and is very close to the centre of our X-ray aperture for the NE
component of A\,3192. \citet{2012ApJ...748L..23H} choose radii of
$r{=}250$ kpc for both apertures which, at $z = 0.425$ and $z =
0.168$, corresponds to 45$''$ and 87$''$, respectively. Their
weak-lensing mass estimate for A\,3192 is $M(<$250 kpc$) \simeq 3
\times 10^{13} M_{\odot}$, in perfect agreement with our X-ray-based
mass estimate of $ 2.80 \times 10^{13} M_{\odot}$.

In order to compare the properties of A\,3192 with those of
\source\ it is useful to do so at roughly the same overdensity
$\Delta$. For a galaxy group like the two components of A\,3192, a
radius of 250 kpc corresponds to an overdensity $\Delta$ of about 2200
(see Eqn.~\ref{eqn:delta}), a value that is reached by \source\ at a
radius of 640 kpc (see Section~\ref{sec:spectral}). The total mass of
\source\ at this radius can be obtained from Fig.~\ref{fig:lens_mass};
we find $M(<$640 kpc$) \simeq 8 \times 10^{14} M_{\odot}$. It follows
that, taken together, the components of A\,3192 comprise at most
one-thirteenth of the mass of \source, where both masses are measured
within $r_{\rm 2200}$.

% ======================================================================%
%                       BIBLIOGRAPHY
% ======================================================================

\bibliographystyle{mn}  
\bibliography{mn-jour,paper0358}

\end{document}